\documentclass[12pt,preprint]{aastex}

\shorttitle{Interaction of YSO Jets}
\shortauthors{Cunningham, Frank, \& Blackman}

\begin{document}

\title{Protostellar Jets and Turbulence in Molecular Clouds: The Role of Interactions}

\author{Andrew J. Cunningham\altaffilmark{1}, Adam Frank\altaffilmark{2},
 Eric G. Blackman\altaffilmark{3}}
\affil{Department of Physics and Astronomy, University of Rochester,
    Rochester, NY 14620}

\altaffiltext{1}{ajc4@pas.rochester.edu}
\altaffiltext{2}{afrank@pas.rochester.edu}
\altaffiltext{3}{blackman@pas.rochester.edu}

\begin{abstract}
We present a series of numerical studies of the interaction of
colliding radiative, hydrodynamic young stellar outflows. We study the
effect of the collision impact parameter on the acceleration of
ambient material and the degree to which the flow is isotropized by
the collision as a mechanism for driving turbulence in the parent
molecular cloud. Our results indicate that the high degrees of
compression of outflow material, achieved through radiative shocks
near the vertex of the interaction, prevents the redirected outflow
from spraying over a large spatial region.  Furthermore, the collision
reduces the redirected outflow's ability to entrain and impart
momentum into the ambient cloud.  Consideration of the probabilities
of outflow collisions leads us to conclude that individual low
velocity fossil outflows are the principle coupling between outflows
and the cloud.
\end{abstract}


\section{Introduction} \label{s1}
Molecular Clouds have long been a subject of interest in astrophysics
since they are the exclusive environments in which stars form in
galaxies.  The expected lifetimes for molecular clouds has become a
topic of considerable debate as numerical simulations have shown that
MHD turbulence, the nominal means of support for the clouds against
self-gravity, will decay on a crossing timescale
\citep{maclow,stone,vazquez-semadeni}.  In light of this result
it is difficult to understand why molecular clouds do not fully
collapse in an efficient burst of star formation on timescales no
longer than a few crossing times. Thus it appears that either
molecular clouds are transient features or they are resupplied with
turbulent energy through some other means.

Jets and molecular outflows are recognized as a ubiquitous phenomena
associated with star formation.  It is expected that most if not all
low mass stars produce a collimated outflow during their formation
from a parent molecular cloud core (massive stars may also produce
collimated outflows though this point remains somewhat speculative
\citep{Shepherd}).  Stars do not, however, form in isolation.  Rich star
forming regions such as Orion can contain as many as 1000 stars per
$\textrm{pc}^3$ \citep{testi}.  Low mass star forming regions such as
Taurus or Perseus will contain hundreds of stars in a similar volume.

The ubiquity and high density of outflows from young stars make them
an intriguing candidate for the source of turbulent energy in
molecular clouds.  The idea that feedback from TT winds could lead to
a self-regulating state of star formation dates back as far as
\cite{Norman Silk}. Consideration of combined energy budget for the
outflows in some clouds compared with the energy in the cloud's
turbulent motions support notions of feedback showing an approximate
balance between outflow input and turbulent support \cite{Bally,Bally
Reip,Knee,Matzner 2002,Warin}. Thus the combined action of many
outflows could, in principle, provide the required deposition of
turbulent energy to support a cloud against collapse. More recent
observational studies have explored multiple (though apparently)
non-interacting outflow structures in individual clouds and come to
similar conclusions. For example direct observational evidence showing
that the giant stellar outflows associated with HH 300 and HH 315 have
disrupted their cloud's density and velocity distributions at parsec
scale distances from their source has been provided by \citet{Arce
2003}. The actual global disruptive effect these individual flows have
depends on the ability that these outflows have to impart their
momentum into their parent clouds by entraining and accelerating
molecular gas \citep{Arce 2003, Arce 2001}.

While invoking jets and outflows to drive turbulent motions appears
attractive for molecular cloud studies, there is a potential problem
with such a scenario. The principle means of energy transfer from jet
to cloud appears to come via shock waves, the so-called ``prompt
entrainment'' mechanism \cite{Chernin}. This is to be compared with
``turbulent entrainment'' mechanism which occurs via a turbulent
boundary at the edge of a jet \cite{Canto}.  Thus the effect of a
single supersonic outflow is bounded by the shock wave which defines
it.  Only those regions of a cloud which have been swept over by the
outflow will gain any energy. Given such a localization of energy and
momenta deposition, the action of multiple, isotropically oriented
outflows is required to drive the random motions associated with
isotropic turbulence.  Somehow the energy and momenta in the localized
region engulfed by a jet or outflow must be randomized and distributed
over many scales.  The results presented here consider the facility of
the collision of two protostellar outflows toward this result.  We
consider the interaction of two heavy jet flows oriented at
$90\,^{\circ}$ to each other with different impact parameters.  Our
goals in this study are to examine the resultant flows and attempt to
distinguish between those with low and high impact parameters in terms
of how ambient gas is accelerated.

\section{Motivation} \label{s2}
We estimate the probability that two protostellar outflows interact as
a function of protostellar density in the cloud.  We consider a volume
$V$ that contains an average outflow density $N$ and assume that each
protostar emits a bipolar outflow.  We approximate the volume of these
bipolar outflows as that of a cylindrical column of length $L$ and
radius $R$.  The outflow fill ratio $\frac{V_{outflow}}{V}=\pi R^2 L
N$ provides an estimate of the volume fraction of the cloud that is
occupied by outflows.  Assuming that the production frequency of
outflows in the cloud is constant we can cast the density of outflows
active at any given instant in terms of the stellar density $N_*$ as
$N=N_* \frac{t_{outflow}}{t_{cloud}}$.  The probability two active
outflows occupy the same region of space in the cloud at the same time
is then $P \sim \left[\frac{V_{outflow}}{V}\right]^2$. Solving for
$N_*$, we have
 \begin{eqnarray}
N_*(P) = \frac{\sqrt{P}}{\pi R^2 L} \frac{t_{cloud}}{t_{outflow}}.
\label{N}
\end{eqnarray}
We define $N_{critical}$ as the protostellar density that achieves a
volume fill ratio of 10\% bowshock overlap $N_{critical} \equiv
N(0.1)$.  Above this intersection probability we expect the effect of
collisions to become appreciable.  We assume values for the typical
protostellar outflow size as $L=1~\textrm{pc}$, bow shock radius
$R=R_{bs}=0.1~\textrm{pc}$, outflow lifetime
$t_{outflow}=2\times10^5~\textrm{yr}$, and cloud lifetime
$t_{cloud}=10^7~\textrm{yr}$ \citep{palla}.  These values yield a
stellar density $N_{critical}=500~\textrm{pc}^{-3}$. This is
comparable to the protostellar density of many star forming regions.
Outflow interactions are therefore statistically likely to occur in a
typical star forming region.

If the collision of outflow streams from adjacent YSO's contribute to
the turbulent energy budget of their parent cloud, it would do so by
increasing the rate at which the flow imparts momentum into the
surrounding molecular gas.  This could occur if the redirected outflow
has a volume greater than the individual outflows.  Also if the
redirected flow generates more ``splatter'', in the sense that a wider
range of scales become energized though vortices generated during the
collision, then the increased rate of momentum deposition into the
ambient molecular gas would result in an increased rate of generation
of turbulent energy and could thereby provide support for the parent
cloud against gravitational collapse and star formation.

Beyond the issue of driving turbulence, the interaction of jets and
outflows is of interest in its own right.  While the propagation of
single jets and outflows has received considerable attention
\citep{Lee 2001, Ostriker} and seems to be fairly well understood, as
is their interaction with clumps \citep{RagaCanto} and side winds
\citep{Lebedev 2004}, the interaction {\it between} such flows is
relatively unexplored.

\section{Computational Method and Initial Conditions}
\subsection{Method: Numerical Code}
In order to explore the efficacy of jet and outflow collisions at
stirring the ambient media we have carried out a series of simulations
using simplified initial conditions.  In our study we have carried
forward hydrodynamic simulations of the interaction of two orthogonal
outflows.  Our simulations include the effect of radiative energy loss
on the flow.  We investigate the role of impact parameter and degrees
of collimation.  Our simulations are carried out in 3D using the
AstroBEAR adaptive mesh refinement (AMR) code. AMR allows high
resolution to be achieved only in those regions which require it due
to the presence of steep gradients in critical quantities such as gas
density. The hydrodynamic version of AstroBEAR has been well tested on
variety of problems in 1, 2, 2.5D \citep{pol,var} and 3D
\citep{Lebedev 2004}.

The BEARCLAW framework on which AstroBEAR is based has been recently
extended to allow efficient parallel computation using MPI on
distributed memory systems by combining two load balancing techniques:
\begin{itemize}
\item{Domain Decomposition: The root level is divided into an
arbitrary number of sub-domains which are balanced across the
processors.}

\item{Dynamic Load Balancing: Each refined grid block is created on
the same processor as its parent grid.  Grids are moved to a lesser
loaded processor at the beginning of each time step if such a move
will reduce the average computational load (measured in CPU time)
across the processors.  This algorithm minimizes inter-processor
communication and overhead and allows efficient load balancing even in
cases where the implicit source term integration requires many
iterations to converge in localized regions of the domain.}
\end{itemize}
Domain decomposition alone cannot achieve efficient load balancing in
the common situation where refined regions are confined to a small
region of the domain.  Dynamic balancing of the grid blocks can fail
in cases where the number of refined grids on any level is less than
the number of available processors.  The combined parallelization
strategy we have employed using both AMR load balancing and domain
decomposition avoids these problems.

\begin{figure}[!h]
\includegraphics[angle=-90,clip=true,width=0.5\textwidth]{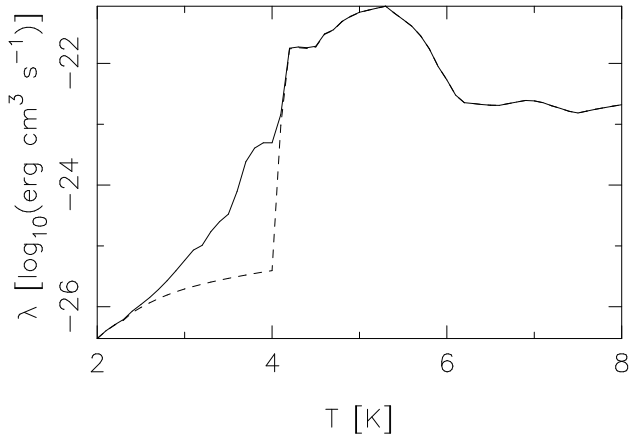}
\caption{The full cooling curve including low temperature contributions
$\lambda$ (solid line) and the atomic line cooling function of
\cite{Dalgarno 1972}(dashed line). \label{cool}}
\end{figure}

For the results presented here, AstroBEAR integrates the system of
equations, $d_t Q + d_x F_x + d_y F_y + d_z F_z = S$.  The vector of
conserved quantities $Q$, the flux function $F$, and the
micro-physical source terms $S$ are given as:
\[
Q=\left[ \begin{array}{c}
\rho \\ \rho v_x \\\rho v_y \\ \rho v_z \\ E \\ \rho_w
\end{array}\right],
F_*=v_* \left[ \begin{array}{c}
\rho \\ \rho v_x \\\rho v_y \\ \rho v_{z} \\ E+P \\ \rho_w
\end{array}\right],
S=\left[ \begin{array}{c} 0 \\ 0 \\ 0 \\ 0 \\ -\left(\frac{\rho}{\mu_H}\right)^2
\lambda \\ 0 \end{array}\right]
\begin{array}{c} \\ \\ \\ \\ \\ , \end{array}
\]
where $\rho$ is the gas density, $v_x$, $v_y$ and $v_z$ are the
Cartesian components of the velocity, $E$ is the total energy, $P$, is
the gas pressure, $\rho_w$ traces the density of injected outflow
material and $\mu_H$ is the atomic weight of hydrogen. The equation of
state used is that of a monotonic ideal gas. While AstroBEAR is
equipped with the capability to track the ionization and chemistry
associated with a number of elements \citep{Cunningham}, we have, for
computational efficacy, elected not to utilize these in the present
calculations due to their high cost.  What matters for our
calculations is only the presence of time-dependent cooling behind
shocks.  As we will see the presence of cooling has a profound effect
on the resulting flow patterns. We do not, however, require a detailed
treatment of the ionization dynamics or chemistry to see these effects
and so we have constructed a cooling function that mimics the effects
of low temperature molecular cooling. This is an alternative to
tracking full non-equilibrium molecular dissociation and cooling.  Our
cooling function (figure \ref{cool}) is given by:
\[
\begin{array}{l}
\lambda(T) = \lambda_{LS}(\chi,T) + \lambda_{DM}(T) \\
\chi(T) = \left\{ \begin{array}{ll}
    0  & \textrm{for}~T\le3000~\textrm{K} \\
    \min\left(\frac{T-3000~K}{7000~\textrm{K}},~0.9\right) & \textrm{for}~3000~\textrm{K} < T < 12600~\textrm{K} \\
    1 & \textrm{otherwise}
\end{array} \right.
\end{array}
\]
where $\lambda_{DM}$ is the atomic line cooling function of
\cite{Dalgarno 1972}, $\lambda_{LS}$ is the molecular cooling function
of \cite{Lepp 1983} for a hydrogen gas of density $10^4~\textrm{cm}^{-3}$ with
a fraction of H atoms in the molecular state $1-\chi$.

The calculations discussed here have been carried out using a spatial
and temporal second order accurate MUSCL scheme using a Roe-average
linearized Riemann solver.  The MUSCL scheme employed here achieves
second order spatial accuracy by performing a MINMOD interpolation of
the primitive fields (density, velocity and pressure) to grid
interfaces.  The TVD-preserving time stepping method of \cite{Shu
1988} is used to advance the solution.  The micro-physical source
terms are handled separately from the hydrodynamic integration using
an operator split approach.  The source term is integrated using an
implicit fourth-order Rosenbrock integration scheme for stiff
ODE's. We have made use of the Local Oscillation Filter method of
\cite{sutherland} using a viscosity parameter $\alpha=0.025$ to
eliminate numerical instabilities that can occur near strongly
radiative shock fronts.

\subsection{Model: Colliding Jet Simulations}
We have carried out a series of six simulations to investigate the
capacity for two interacting outflows to excite motions on a range of
scales and the efficacy with which they accelerate ambient gas.  The
parameters that are common to all of the simulations are given in
table \ref{paramtable}. Our computations are carried forward with a
base grid that resolves the jet inflow with two cells per jet radius.
Three levels of AMR refinement, each with a refinement ratio of two,
produce and effective resolution of 16 computational cells per jet
radius.  The flows in these simulations are characterized by strong
cooling, typical for YSO outflows.  We characterize the strength of
the cooling by defining a cooling parameter $\sigma$ which is the
ratio of $\tau_{cool}$, the time for post shock ambient material to
cool to $10^4~\textrm{K}$, to $\tau_{cross}$,the time for post shock
ambient material to cross $r_j$.  Typical values for YSO jets and
outflows are such that flows with $\sigma << 1$ are expected.  For the
parameters chosen here, the cooling parameter is given by $\sigma =
\tau_{cool}/\tau_{cross} \approx 0.01$.

Each simulation consists of two identical orthogonal jets.  The jets
are launched into domain via fixed cells embedded within the boundary.
In the fixed cells the outflows maintain a uniform density and
temperature. We use a velocity distribution characterized by a shear
parameter $s$ and spray angle $\psi$:
\[
\begin{array}{l}
\alpha(r)=1-(1-s)\left(\frac{r}{r_j}\right)^2 \\
v_\parallel(r)=v_j\alpha(r)\cos(\frac{r}{r_j} \psi) \\
v_\perp(r)=v_j\alpha(r)\sin(\frac{r}{r_j} \psi)
\end{array}
\]
where $r$ is the distance from the center of the jet axis.  Except
where the jets are launched into the domain, the boundary conditions
used are extrapolation.  To prevent the expansion of the jet inflow
boundary with time, a ring of zero velocity that extends to $1.125r_j$
is maintained around the jet launching region.  This is done to keep
the jet conditions from bleeding into the extrapolation boundary zones
surrounding the jet.

\begin{table}[!h] \caption{Simulation Parameters. See text for details. \label{paramtable}}
 \begin{tabular}{l l}
 \tableline
 Jet Radius $r_j$ & $100~\textrm{AU}$ \\
 Computational cells per $r_j$ & $16$ \\
 Jet Density $\rho_a$ & $7500~\textrm{cm}^{-3}$ \\
 Jet Peak Velocity $v_\circ$ & $200~\textrm{km s}^{-1}$ \\
 Jet Temperature & $10^4~\textrm{K}$ \\
 Ambient Density $\rho_a$ & $2500~\textrm{cm}^{-3}$ \\
 Ambient Temperature $T_a$ & $200~\textrm{K}$ \\
 Shear parameter $s$ & 0.9 \\
 \tableline
 \end{tabular}
\end{table}

We have carried out two sets of simulations.  Each set is defined in
terms of the opening angle of the outflow, $0^\circ$ and $15^\circ$.
Hereafter we will use the term ``jets'' to refer to cases with
$0^\circ$ spray angle and the term ``wide angle jets'' (WAJ) to refer
to the cases with $15^\circ$ spray angle. Note that this is the half
opening angle of the outflow measured from the jet axis to limit of
the beam.  Each simulation set has been carried out over the impact
parameters $b=0$, $b=r_j$, and with $b$ sufficiently large that the
bow shocks do not intersect.  We use the results of the
``non-interacting'' simulations as a control case to contrast with the
effects of the interacting winds at smaller impact parameters.  For
the collimated jet non-interacting case $b=5.33r_j$ and the WAJ case
$b=8r_j$.

We note that due to computational cost we were only able to follow the
evolution of flows for time and length scales that are short compared
to actual YSO environments.  This is often the case for YSO jet
simulation studies as the expectation is that scalings allow the
behavior present in the simulation to characterize what occurs in YSO
environments.

\section{Results}
\subsection{Morphology}
\begin{figure}[!h]
\includegraphics[clip=true,width=0.32\textwidth]{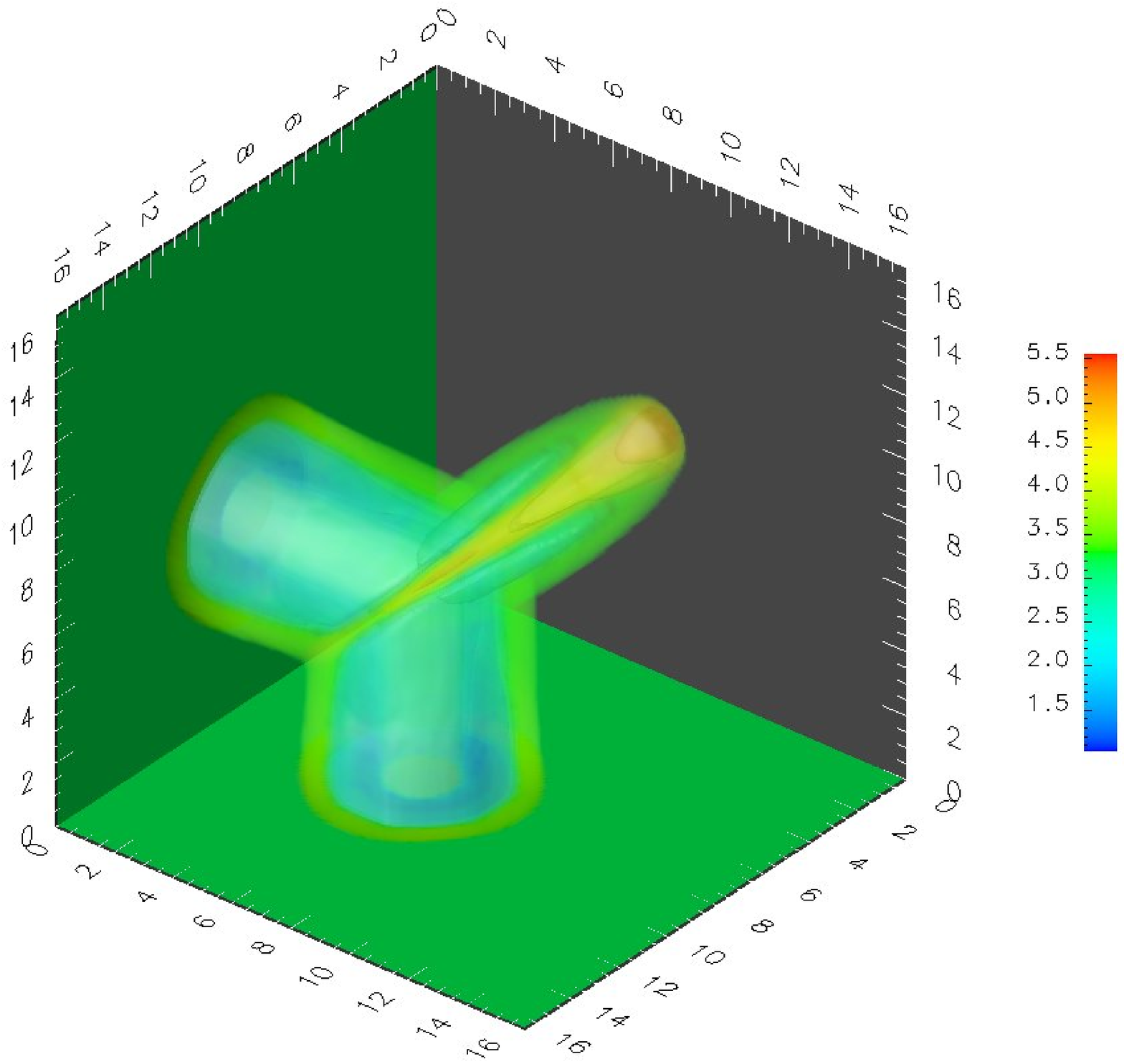}
\includegraphics[clip=true,width=0.32\textwidth]{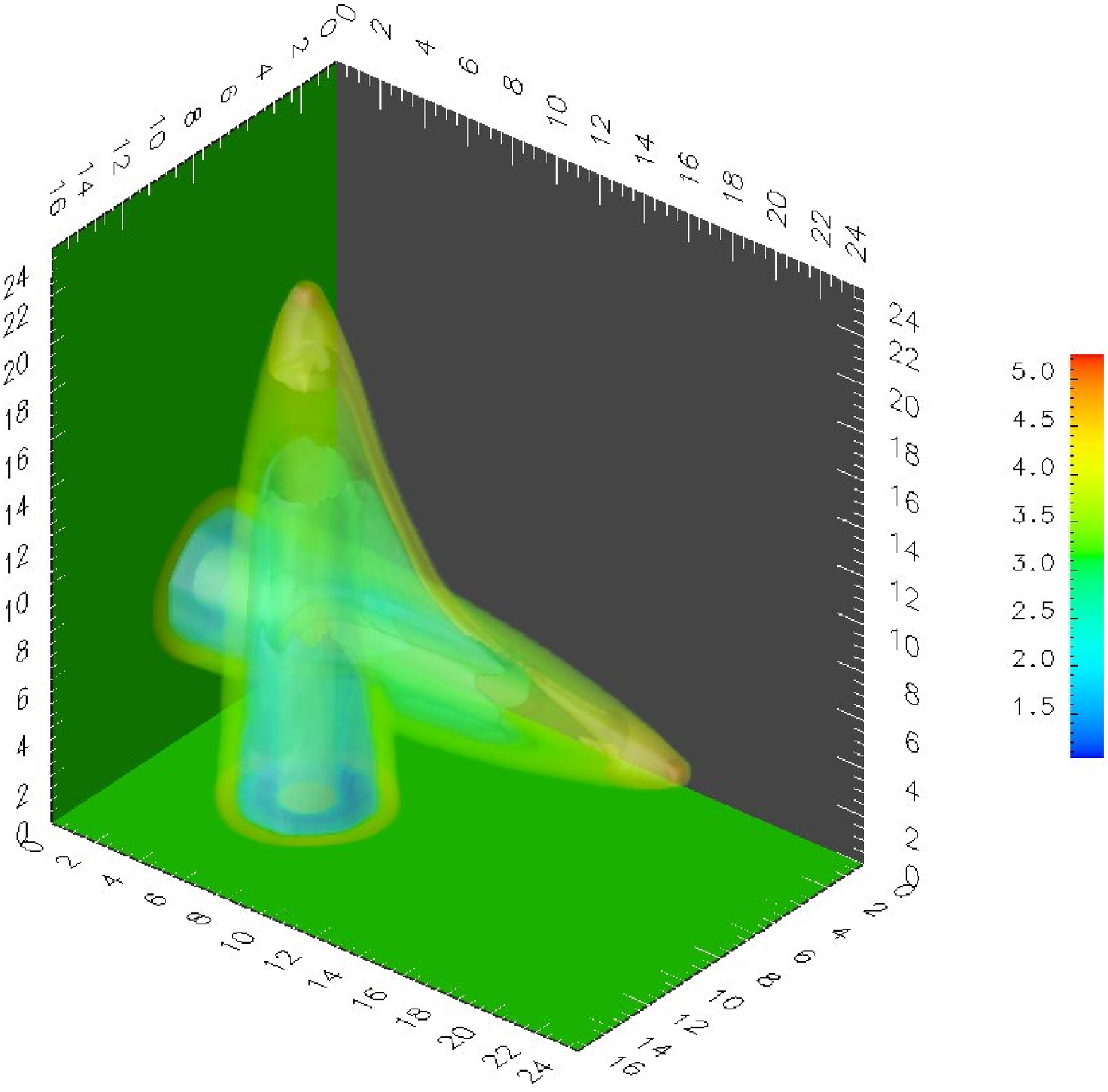}
\includegraphics[clip=true,width=0.32\textwidth]{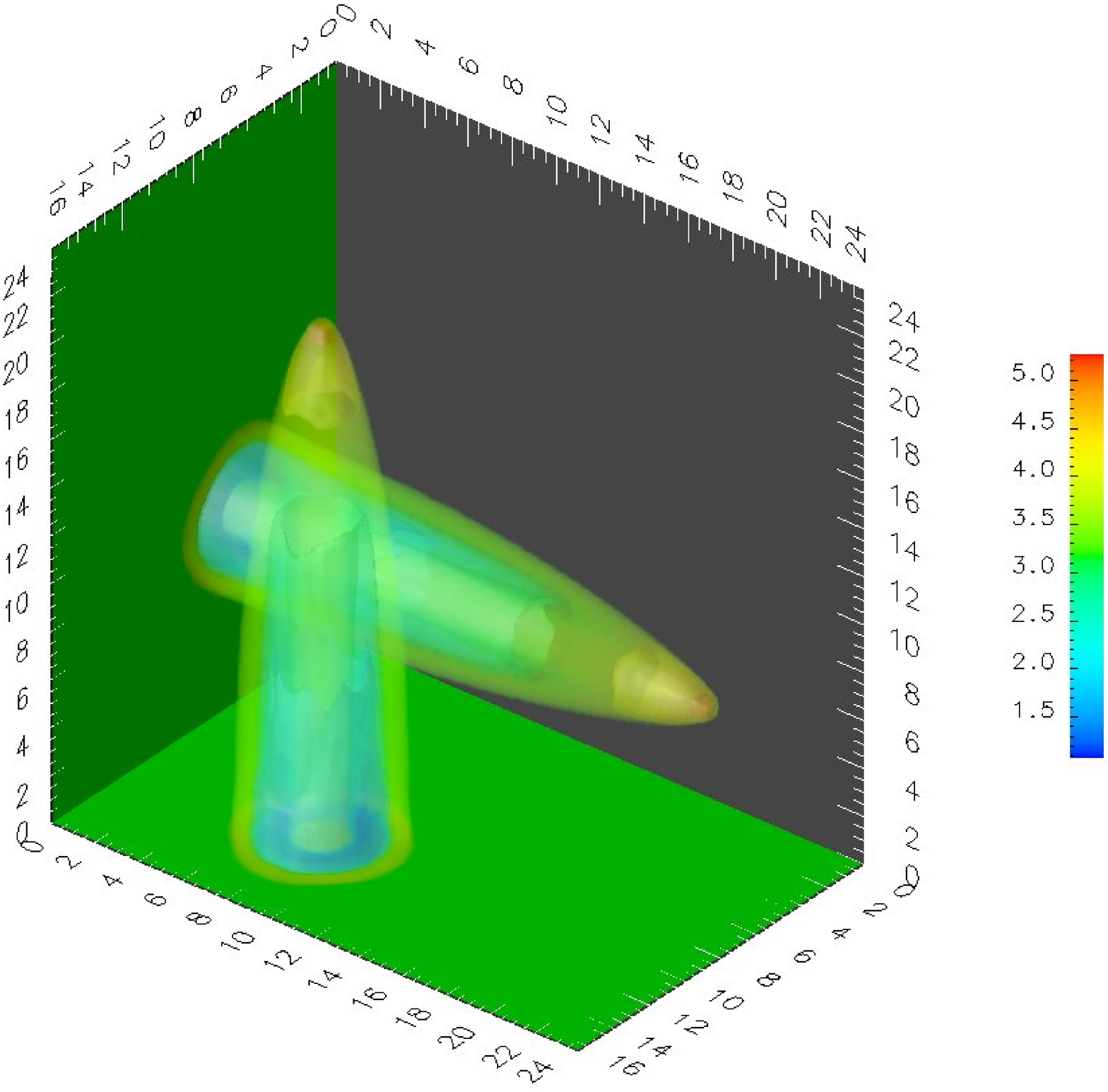}
\caption{A semi-transparent volume rendering of gas density in units
of $\log(\textrm{cm}^{-1})$ at time $t=75~\textrm{yr}$ for the collimated jet
simulations.  Semi-transparent isosurfaces are plotted at
$\log(\rho)=2.0, 3.0, 4.0, 5.0~\textrm{cm}^{-1}$.  The axes are
labeled in units of $r_j=100~\textrm{AU}$.  Impact parameters of $b=0$
(left), $b=2r_j$ (center), and $b=5.33r_j$ (right) are shown.
\label{f2}}
\end{figure}

Figure \ref{f2} shows a semi-transparent volume rendering of the
logarithm of gas density of the jet collision simulations for each
impact parameter studied.  Isosurface contours are shown at $\log
\left[ \rho \left(\textrm{cm}^{-3} \right) \right]=2.0,
3.0,4.0,\textrm{and }5.0$.  When the directly colliding jets are
perfectly aligned we observe a complete redirection of the flow.  The
two jets merge into a single beam.  The redirection occurs via the
formation of oblique shocks within each beam at the point of impact.
Simple momentum conservation considerations show that redirection
angle, $\tan \theta=\left(\frac{\rho_2 v_2}{\rho_1 v_1} \right)$ where
$\rho_1$ and $v_1$ are the density and velocity of the initial jet in
the $\hat{x}$ direction and $\rho_2$ and $v_2$ are the that of the
initial jet in the $\hat{y}$ direction.  Because of the identical
conditions in each of the orthogonal jets considered here, the
resulting flow exits at an angle of $\theta = 45^o$ from either jet
axis.  Note that the resultant jet beam emanating from the point of
impact has a small opening angle and hence its radius throughout the
computational space is smaller than that of the original jets.  We
find that this result is strongly dependent on radiative cooling.
When the cooling is turned off completely a broader redirected flow is
obtained.  This result is to be expected and similar behavior was
obtained in the study of conical converging flows studied by
\cite{Cantoetal88}.  In particular that study derived relationships
between the converging conical flow of half-angle of incidence, $i$,
and the half-opening angle, $\alpha$, of the resultant flow.  The
resultant half-opening angle, it was found, depends on the inverse
compression ratio behind the conical shock, $\zeta = \rho_u/\rho_d$,
where subscripts u and d refer to upstream and downstream conditions.

\begin{equation}
\tan\alpha =  \frac{(1-\zeta)-\sqrt{(1-\zeta)^2 - 4\zeta
\tan^2 i}}  {2\tan i} \label{alpha}
\end{equation}

The equation above demonstrates that as the compression ratio
increases (and $\zeta$ decreases), the opening angle of the conical
axial shock will decrease.  This result also applies to the opening
angle in the plane of symmetry of the resultant flow emerging from the
direct interaction of jet outflows.  As $\alpha$ decreases, the radius
of the emerging jet at any point downstream drops as well. The
post-shock compression depends on the degree of radiative cooling.
Thus equation \ref{alpha} predicts that the secondary outflow forming
from collisions of incident streams will become more narrow with more
effective cooling. The shock bounded slab at the vertex of the
interaction region of the simulations presented here achieves $\approx
24\times$ compression.  The predicted half opening angle is then
$\alpha \approx 2.6^\circ$.  The simulations (figure \ref{f4}) reveal
a half-opening angle $\approx 2.7^\circ$, consistent with the analytic
prediction.

While the $b=0$ case produces strong modification of the jet flow, the
grazing collision case, $b=2r_j$, has far less dramatic morphological
consequences. Consideration of figure 2 shows that no global
redirection of material occurs in the $b=2r_j$ case and only a minor
``splash'' of slowly moving post-shock material is dragged away from
the bow shocks surrounding the two jet beams.  Comparison of the
$b=2r_j$ case with the control case shows little morphological
difference. Thus, in terms of modifying the global flow for narrowly
collimated jets with strong cooling, impact parameters of $b<2r_j$ are
required to significantly alter the morphology of the system.

\begin{figure}[!h]
\includegraphics[clip=true,width=0.32\textwidth]{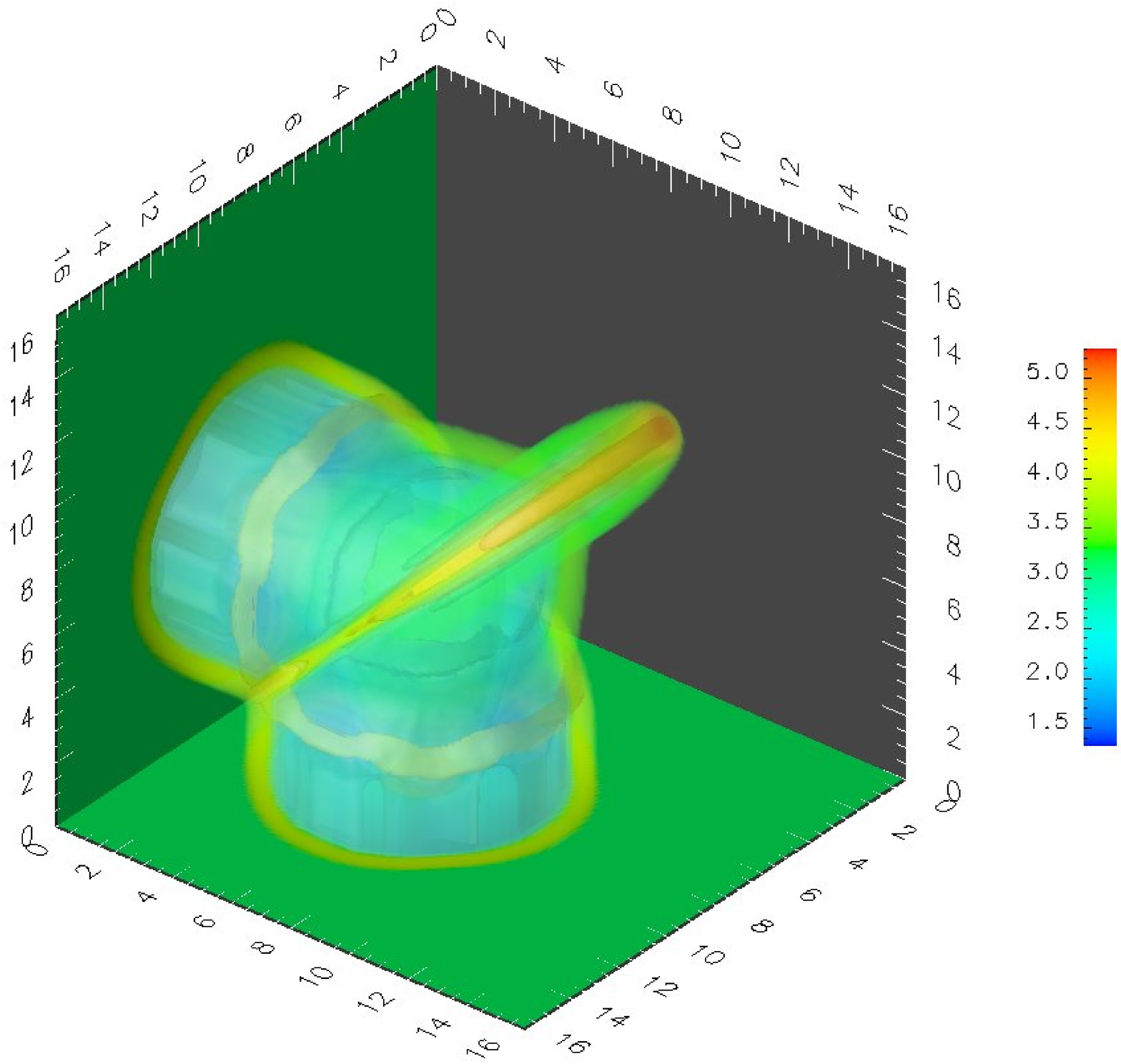}
\includegraphics[clip=true,width=0.32\textwidth]{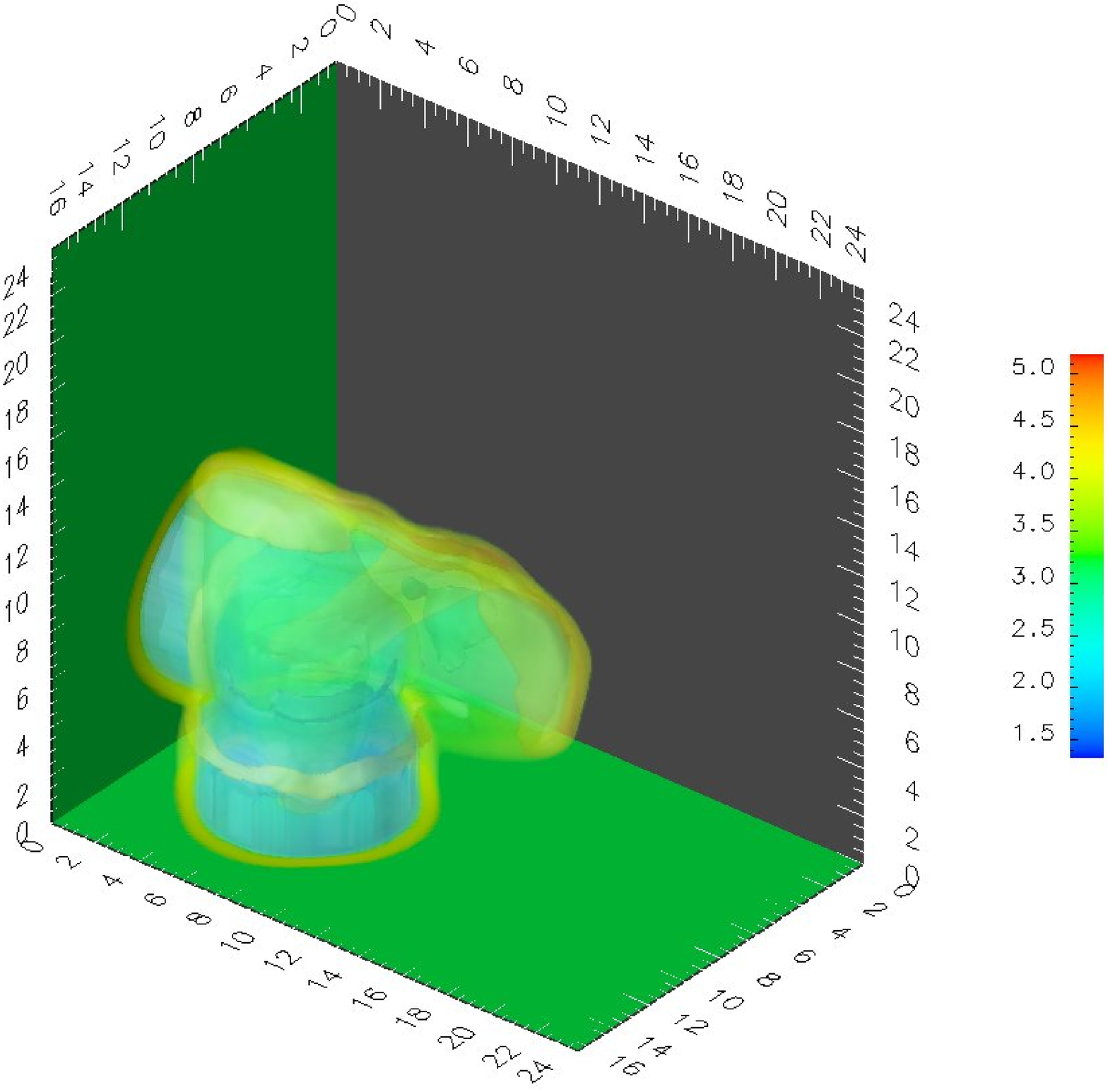}
\includegraphics[clip=true,width=0.32\textwidth]{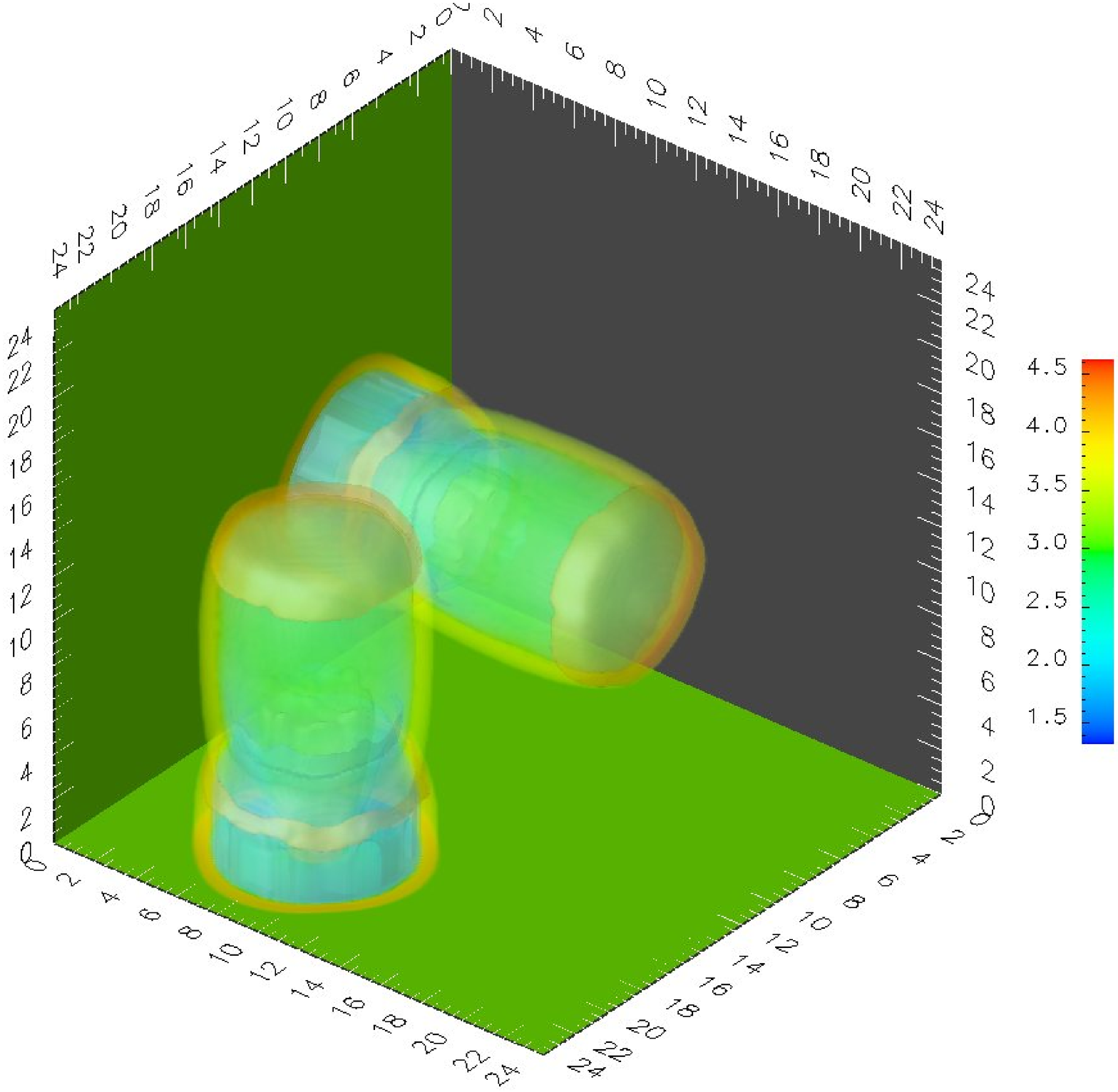}
\caption{A semi-transparent volume rendering of gas density in units
of $\log(\textrm{cm}^{-1})$ at time $t=75~\textrm{yr}$ is shown for
the WAJ simulations.  Semi-transparent isosurfaces are plotted at
$\log(\rho)=2.0, 3.0, 4.0, 5.0~\textrm{cm}^{-1}$.  The axes are
labeled in units of $r_j=100~\textrm{AU}$.  Impact parameters of $b=0$
(left), $b=2r_j$ (center), and $b=8r_j$ (right) are shown. \label{f3}}
\end{figure}

\begin{figure}[!h]
\includegraphics[angle=-90,clip=true,width=0.24\textwidth]{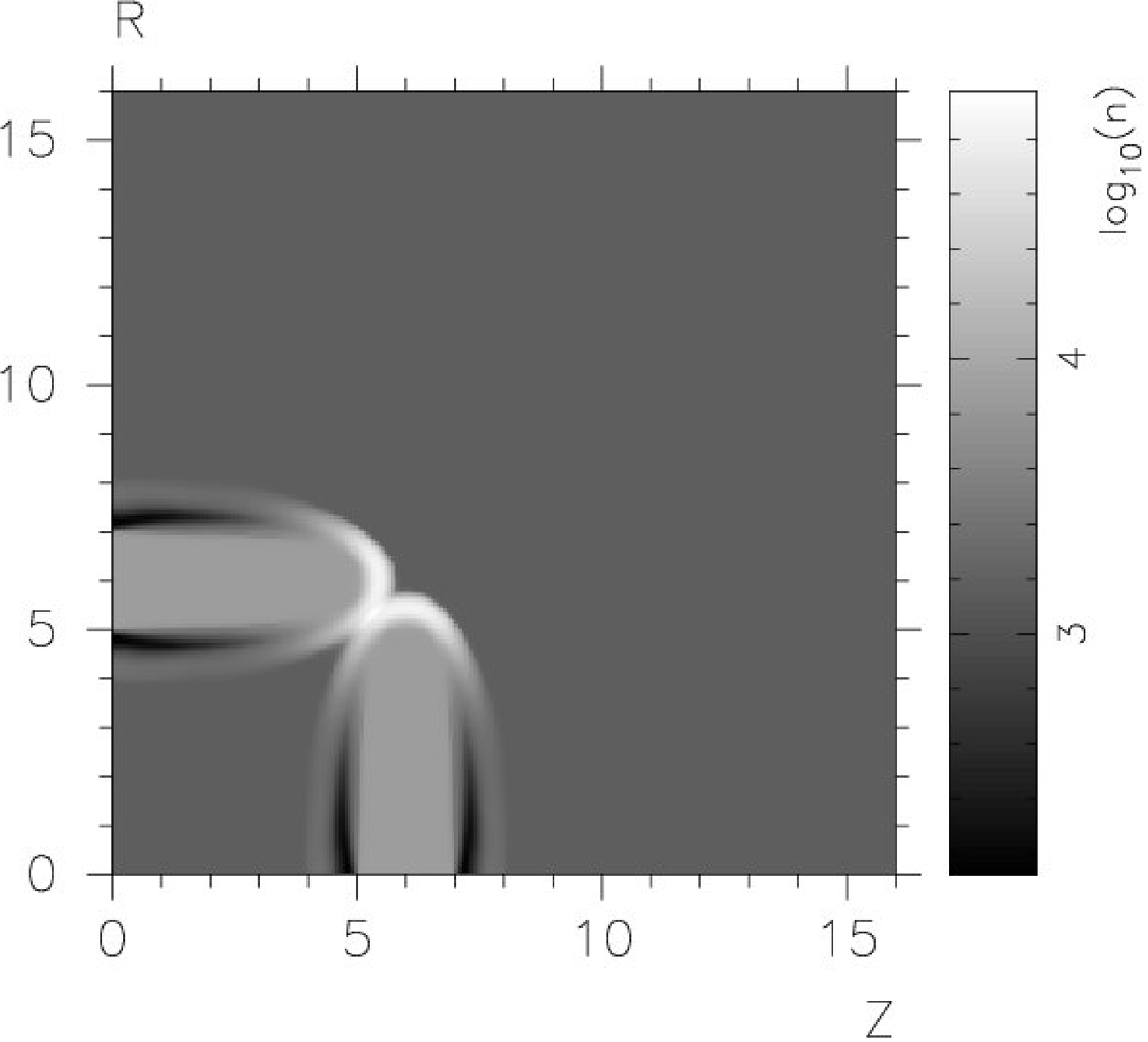}
\includegraphics[angle=-90,clip=true,width=0.24\textwidth]{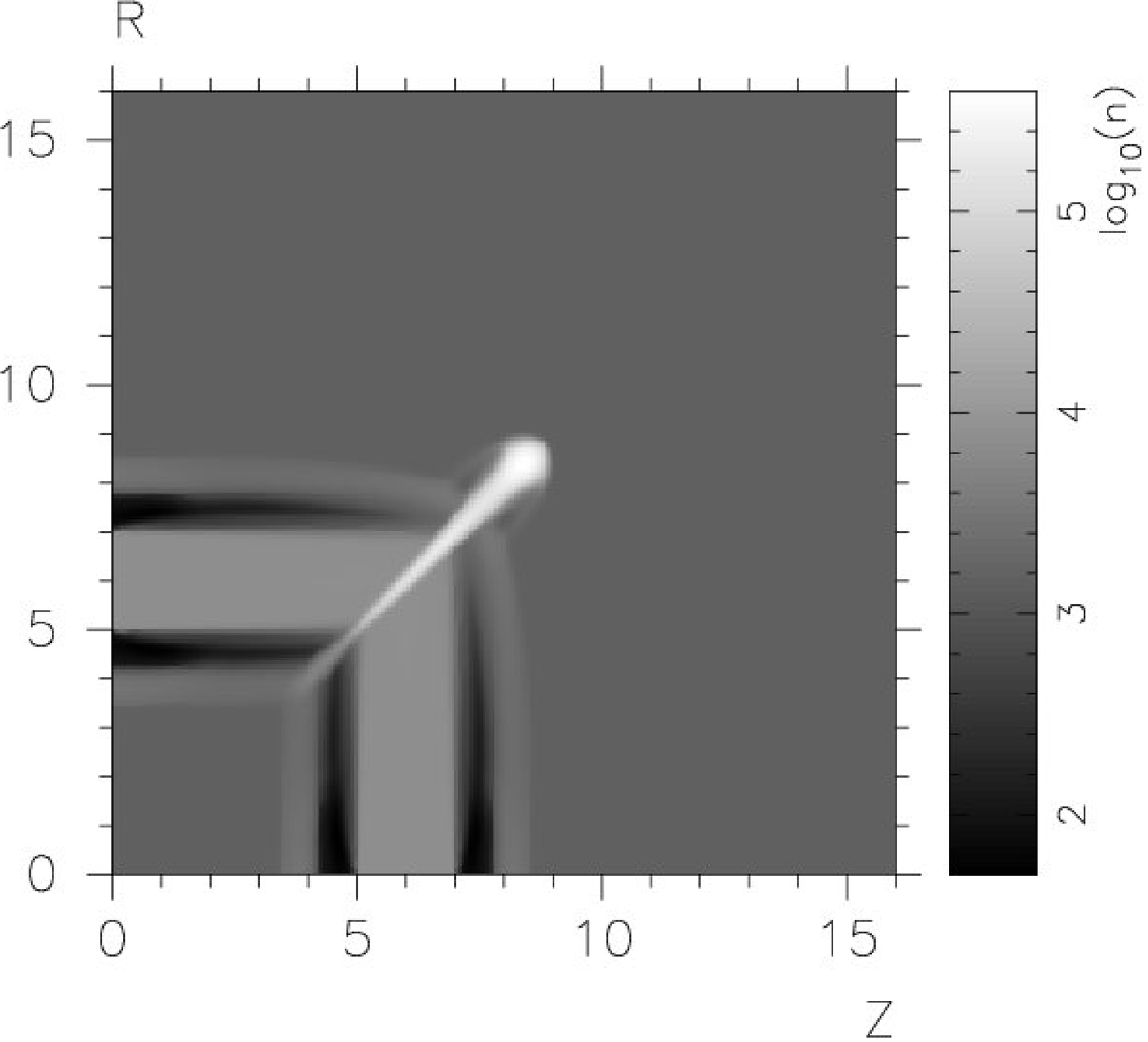}
\includegraphics[angle=-90,clip=true,width=0.24\textwidth]{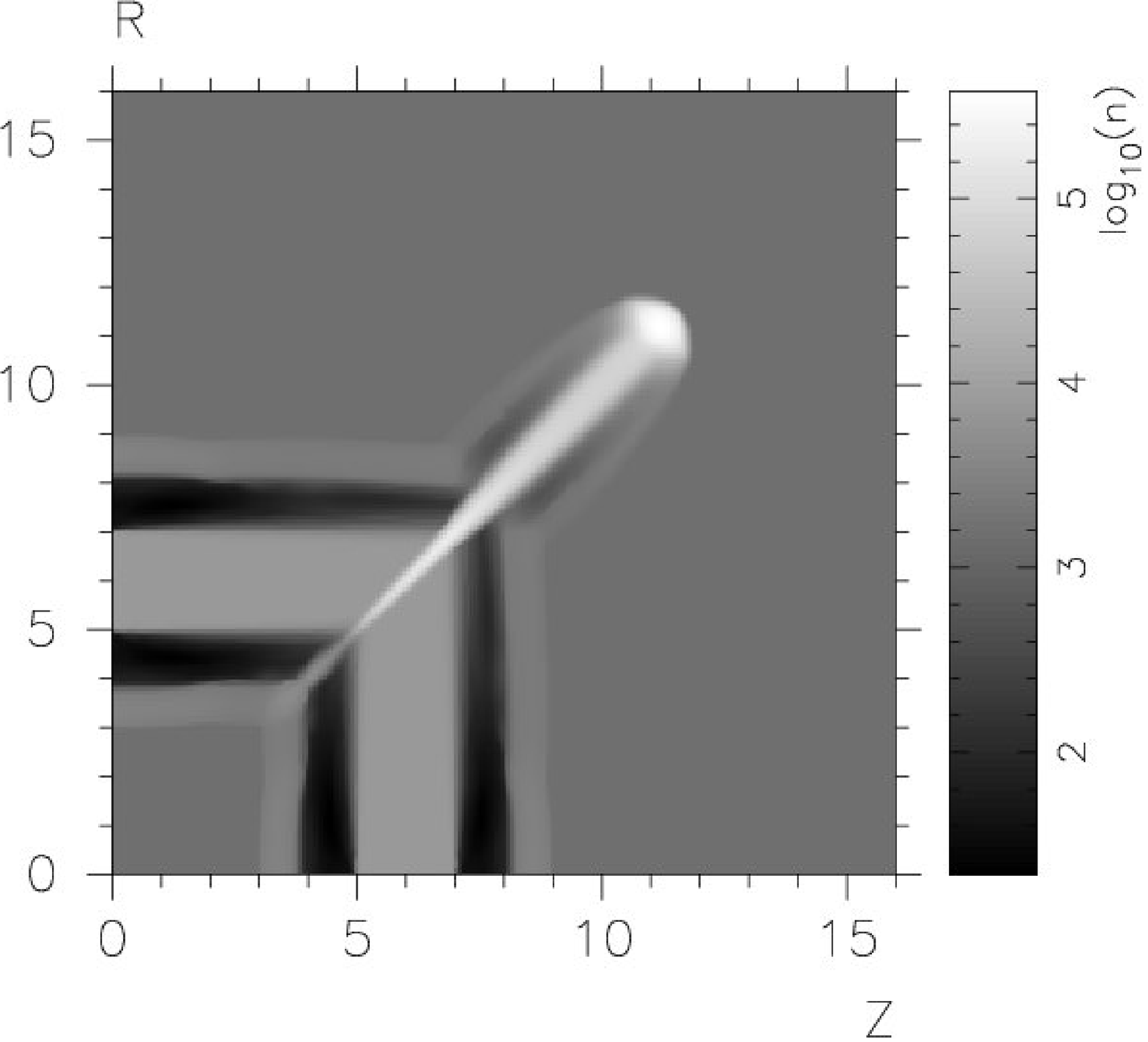}
\includegraphics[angle=-90,clip=true,width=0.24\textwidth]{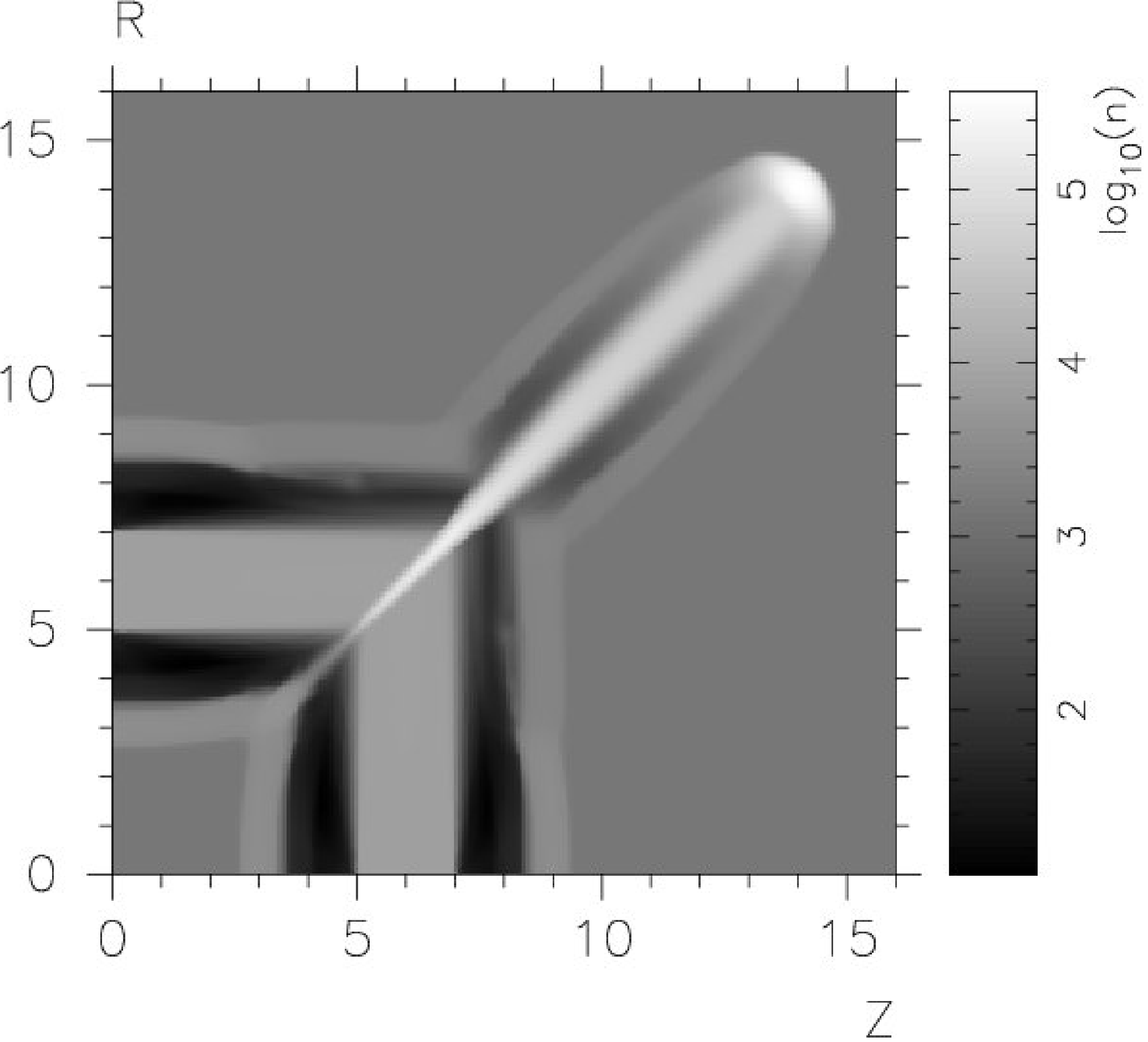}
\caption{Cross sections about the plane of symmetry for the collimated
jet direct collision ($b=0$) simulation of gas density in units of
$\log(\textrm{cm}^{-1})$ are shown at time $t~=18.75, 37.5, 56.25,
75\textrm{yr}$ from left to right.  The axes are labeled in units of
$r_j=100~\textrm{AU}$.
\label{f4}}
\includegraphics[angle=-90,clip=true,width=0.24\textwidth]{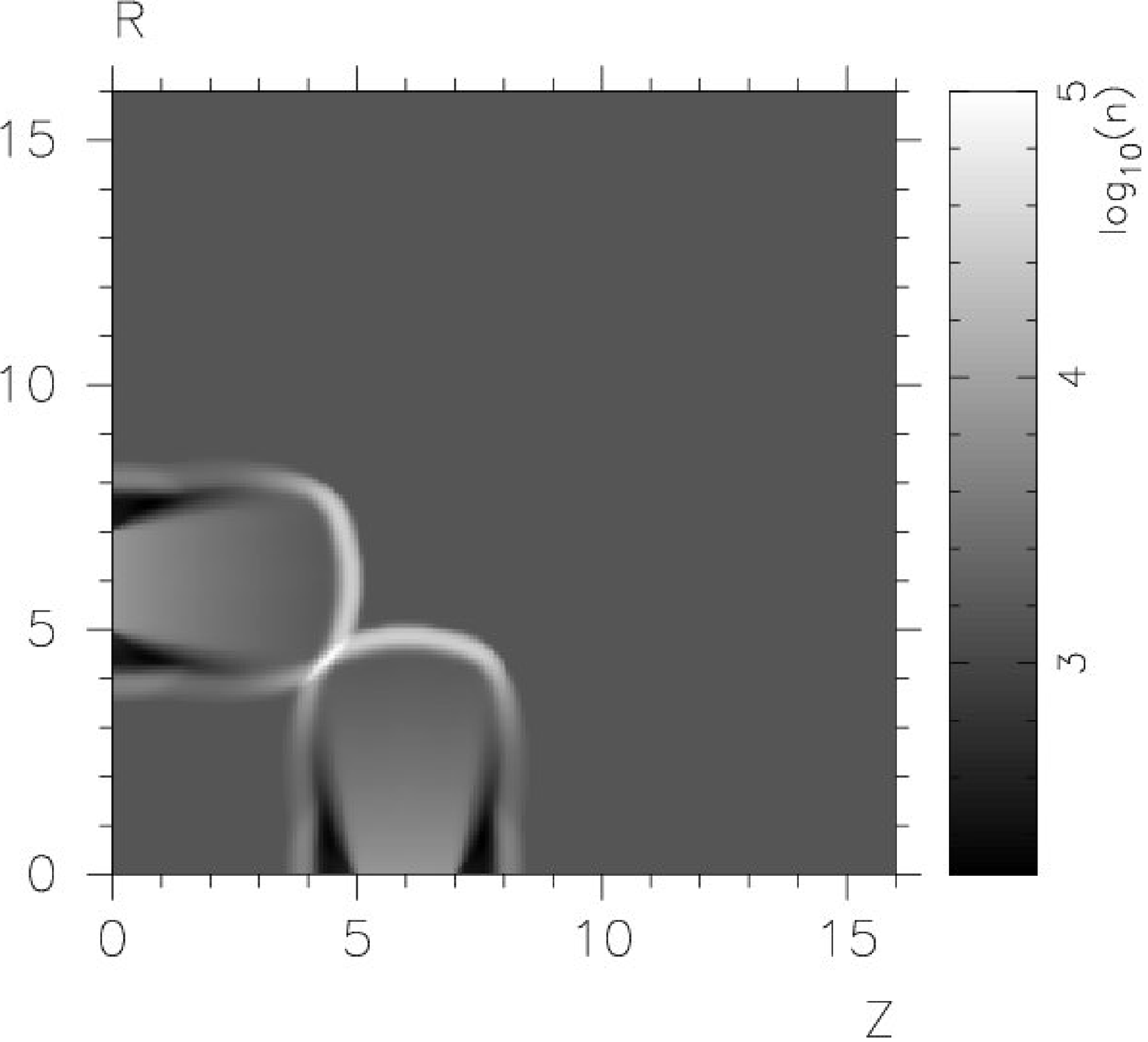}
\includegraphics[angle=-90,clip=true,width=0.24\textwidth]{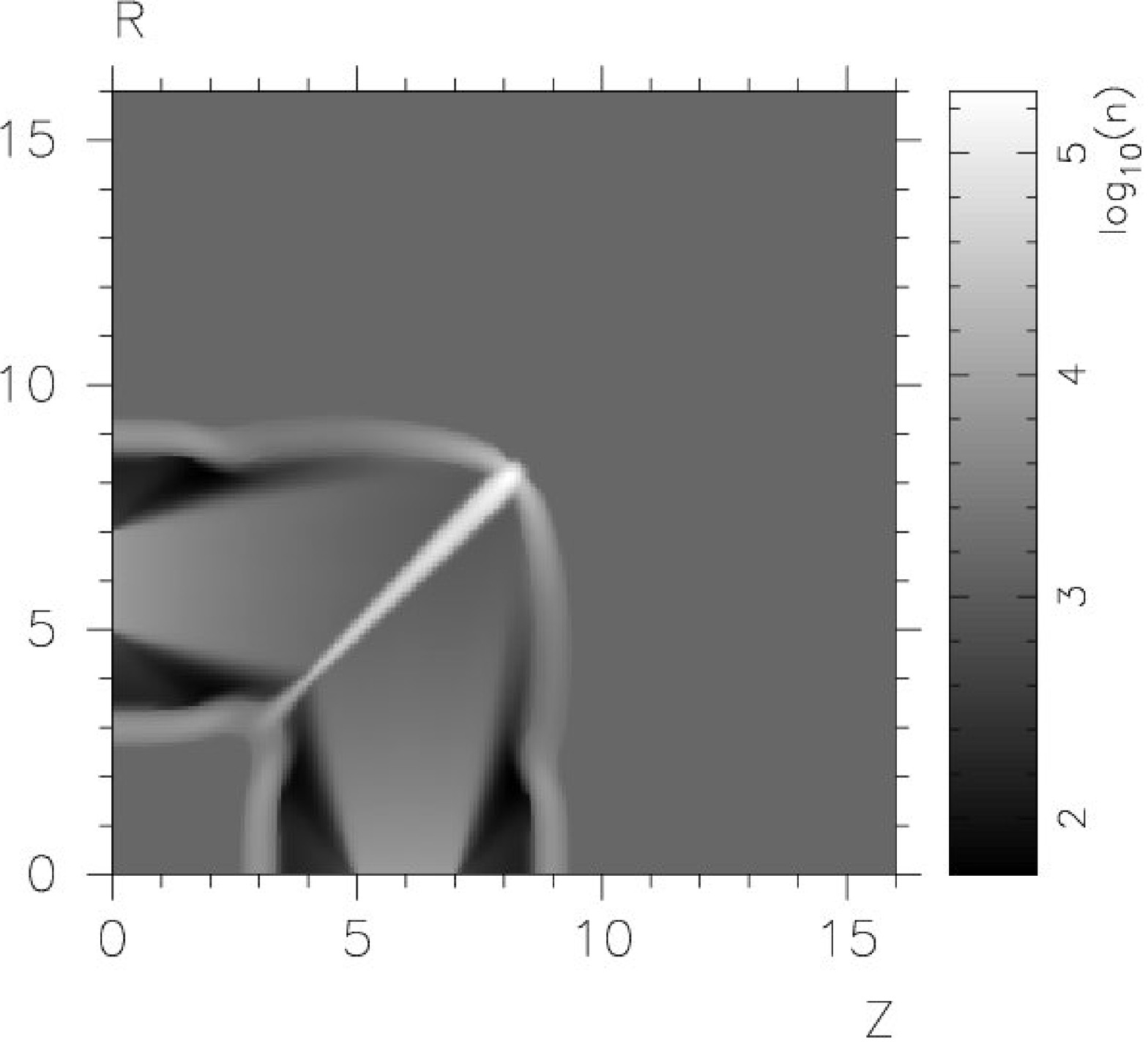}
\includegraphics[angle=-90,clip=true,width=0.24\textwidth]{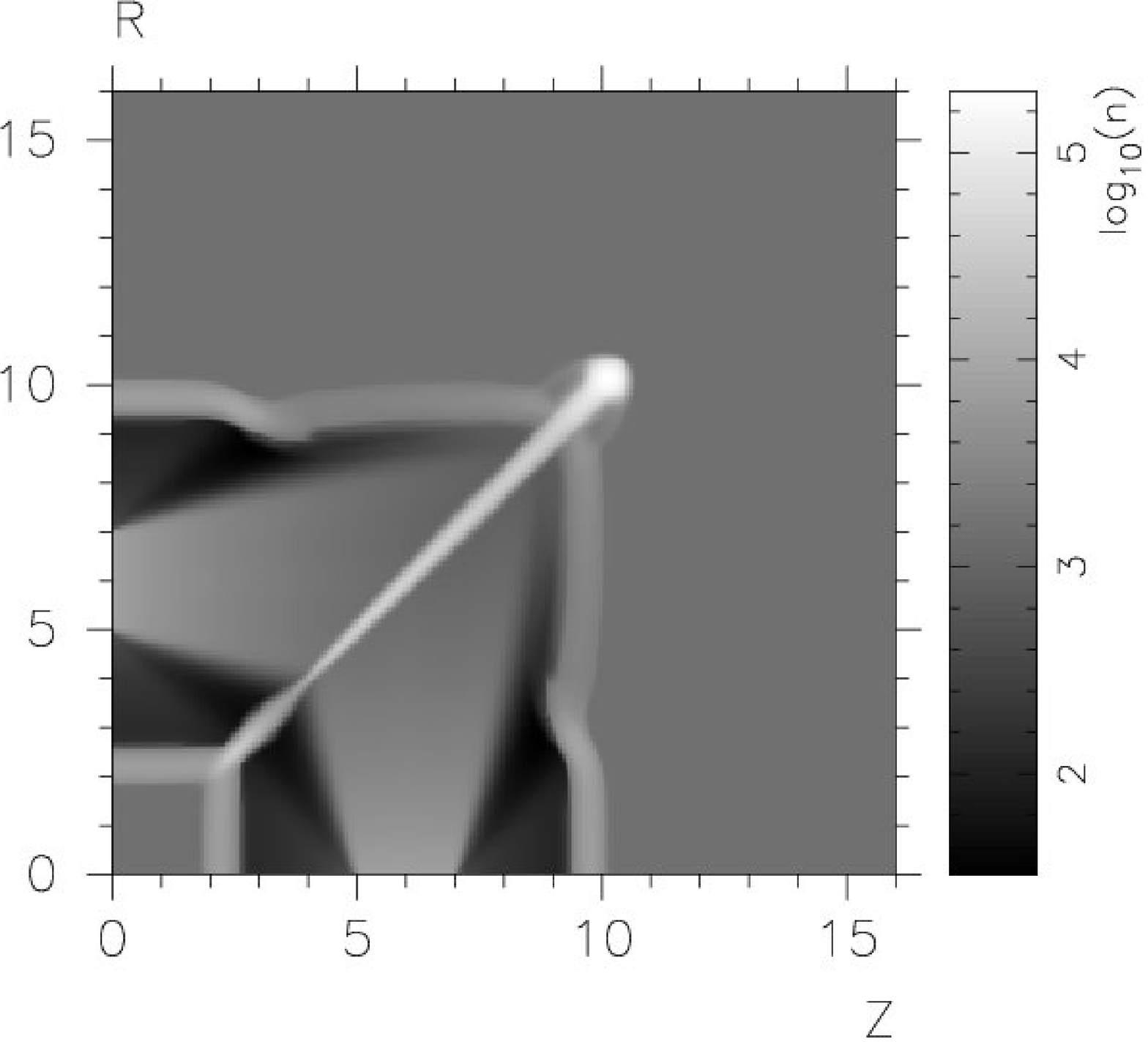}
\includegraphics[angle=-90,clip=true,width=0.24\textwidth]{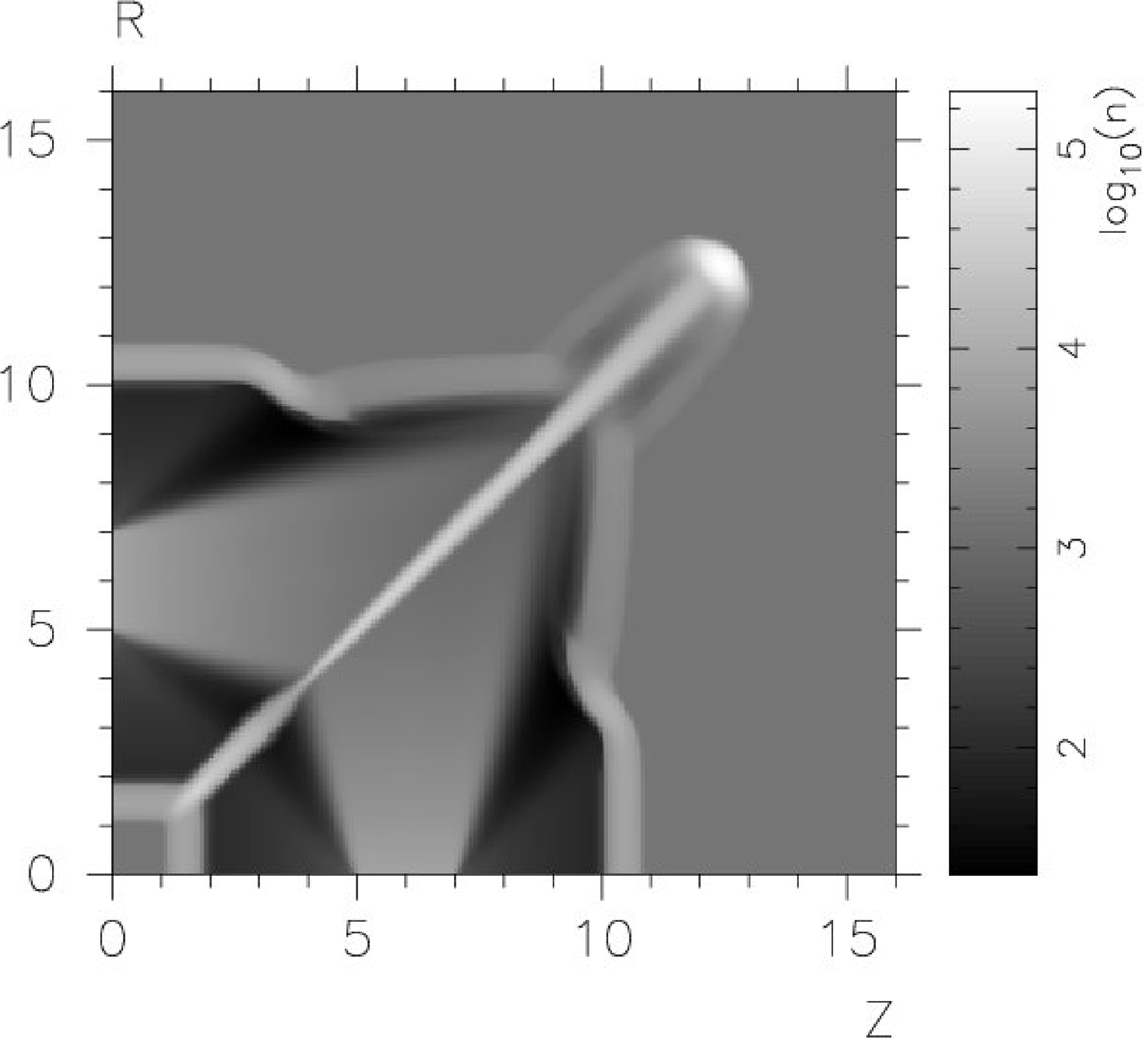}
\caption{Cross sections about the plane of symmetry for the WAJ direct
collision ($b=0$) simulation of gas density in units of
$\log(\textrm{cm}^{-1})$ are shown at time $t=18.75, 37.5, 56.25,
75~\textrm{yr}$ from left to right.  The axes are labeled in units of
$r_j=100~\textrm{AU}$. \label{f5}}
\end{figure}

Figure \ref{f3} presents the results for the WAJ simulations for each
impact parameter. Figure \ref{f3} uses the same visualization
technique described for the collimated jet case.  Note that the
leading edge of the WAJ outflows are driven by an increasingly diluted
wind as the flow evolves due to geometrical effects.  As in the
collimated jet case we see that a direct collision ($b=0$) yields a
compete redirection of the flow.  Once again we see a new, merged flow
extending from the point of impact of the two incident beams. As in
the fully collimated jet collision we also see a resulting flow with a
smaller radii than that for the incident beams.  A significant
difference exists between the collimated and WAJ case however in the
nature of the internal dynamics of the collision.  The collision of
the two flows always creates a shock bounded sheet of gas which,
because of the strong cooling, flows almost parallel to the shock
tangents.  In the study of colliding flows the region between the
shocks is often called a cold dense layer (CDL) \citep{walder}. The
WAJ collision leads to the creation of broader CDL than in the
collimated jet case.  This is simply due to the larger surface area
associated with a conic section through the cone of a WAJ as compared
with that for the cylindrical beam of the collimated jet.  Cross cuts
about the symmetry plane for the $b=0$ collimated jet (figure
\ref{f4}) and WAJ (figure \ref{f5}) cases show that the density of the
redirected flow in the CDL is greatly enhanced across the shock, as
expected.

Consideration of the larger impact parameter simulation shows an
additional significant difference between the collimated and WAJ
simulations.  The wide spray angle ($15^\circ$) of the WAJ simulation
results in strong interaction of the injected beams in the $b=2r_j$
case.  Figure 3 shows that material in the beam of each WAJ is caught
up in the collision.  A more detailed examination of the
results shows that nearly half of the material injected by each WAJ
participates in the interaction.

\subsection{Entrainment of Ambient Gas and Turbulence}
In this section we attempt to extract more quantitative measures of
the interacting jet's ability to excite motions of ambient
material.  One of the most frequently used diagnostics for ambient
mass entrainment are mass-velocity relations.  Molecular outflows
associated with YSOs often exhibit power law mass distributions as a
function of velocity, $M(V) = V^{\gamma}$, at low and intermediate flow
velocities.  A break in the power-law is often present at the highest
velocities associated with the speed of the outflow driver.  As noted
in the introduction it remains unclear if the driver is a
protostellar jet or wide angle wind though the driver speed is
usually close to the escape velocity of the protostar ($V \sim V_{esc}
\sim 100~\textrm{km s}^{-1}$).

\begin{figure}[!h]
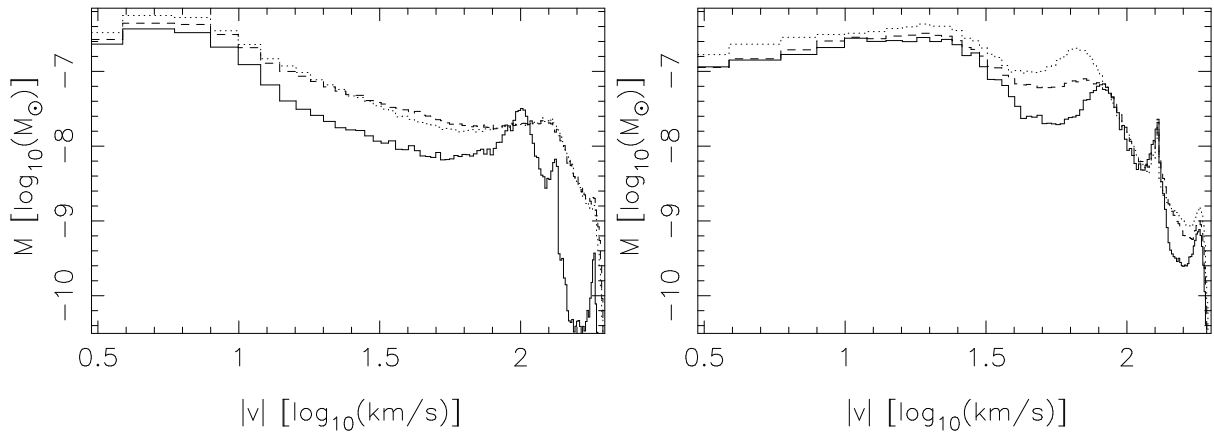

\includegraphics[angle=-90,clip=true,width=0.48\textwidth]{f6a.ps}
\includegraphics[angle=-90,clip=true,width=0.48\textwidth]{f6b.ps}
\caption{Ambient gas mass vs. velocity magnitude power law histograms at
$t=75~\textrm{yr}$ are for the collimated jet (left) and WAJ (right)
simulations.  Solid lines correspond to direct collision, dashed lines
correspond to impact parameter $b=2r_j$ and dotted lines correspond to
the non-interacting case. \label{f6}}
\end{figure}

In Figure \ref{f6} we show the $M \textrm{vs.}~|V|$ plots extracted
from our simulations.  The $M \textrm{vs.}~|V|$ data is extracted from
the simulations by binning the gas in the computational domain into
$2~\textrm{km s}^{-1}$ wide channels. The distribution of ambient mass
is displayed in histogram form as a function of velocity for each
impact parameter in separate plots for the collimated jet and WAJ
cases.  It is clear from the figures that in the collimated jet cases
the interaction does not disrupt the power law behavior of these flows
at intermediate velocities.  The collimated outflow simulations
produce good fits to a power law at intermediate velocities with an
index of $\gamma \sim -1.7$.  This is value is consistent with both
observational \citep{Lada 1996} and theoretical \citep{Smith 1997, Lee
2001} studies which find a range of $\gamma \approx -1.3$ to $-2.5$
for both YSO jet flows and wide angle winds at later stages in their
evolution.  Note that wide angle winds can produce these power law
distributions only in cases in which the ambient medium shows a strong
toroidal density distribution and/or the wind shows significant
asphericity in terms of its pole to equator momentum distribution
\cite{li,Gardiner}. In the WAJ case we do not see such clean power law
behavior at lower velocities.  We attribute this difference to the
fact that we are not using a wide angle wind in the sense discussed
above. The momentum distribution in the jet is not a function of polar
angle and the ambient density is constant.

The most important result in the present context, however, is the fact
that we see no significant differences, in either the collimated jet
or WAJ cases, between simulations with different impact parameters.
The outflow collisions have not greatly enhanced the amount of ambient
material entrained into the flow, nor has it accelerated a significant
fraction of ambient material to high velocity.  The interacting cases
have accelerated only a small fraction of the outflow material to $v
\sim 100~\textrm{km s}^{-1}$ when compared to the non-interacting
control case.  The shape of the $M(v)$ curves at velocities $v <
60~\textrm{km s}^{-1}$ in all cases are similar.  Only at higher
velocities do we see any difference between the various impact
parameters. For the direct collision collimated jet simulation
we find a local maximum in mass occurring at $V \sim~100~\textrm{km
s}^{-1}$ due to the redirected jet material.  The local maximum
corresponding to redirected outflow material in the WAJ $b=0$ and
$b=2r_j$ cases occurs at $V \sim 125~\textrm{km s}^{-1}$.

The excitation of motions of a variety of scales must occur if
outflows are to act as the sources of turbulent energy for the parent
cloud.  It might have been expected that the collision of high speed
outflows would be effective at entraining more ambient material into
the flow while isotropizing the momentum of the jets.  This could
happen if collisions increased the efficiency and rate of momentum
exchange with the environment, relative to the non-interacting case,
by redirecting post-collision material into a wide spray which
generated motions on a variety of scales.  Our simulations show no
evidence that jet collisions with radiative losses create a wide spray
or accelerate more material than in the non-interacting case.
Excluding the local maxima noted previously, the direct collision
cases have entrained less ambient material at all velocities compared
to the non-interacting and weakly interacting cases (figure \ref{f6}).
We attribute this to the reduced bow shock surface area where ambient
material is entrained into the flow. The strong cooling present behind
the collision shock allows the redirected outflowing material to
condense into a thin column (figure \ref{f5}).  The net surface area
of this column is less than what is realized in the absence of the
collision.  The strong dissipational nature of the direct collision
means that we have effectively taken two jets with the associated
capacity to entrain ambient gas and turned them into a single denser,
more narrow outflow.

\subsection{Enstrophy Generation}
The route to turbulence is expected to take the form of an excitation
of modes on a variety of scales.  In the classical theory of
incompressible turbulence \citep{kolmogorov} these modes are vortices
of different eddy size.  A self-similar cascade of vortical motions
from the injection scale down to the dissipation scale is expected to
be the final steady state of sustained turbulence. Thus we expect to
see enhanced enstrophy within the computational domain as a precursor
to the development of turbulence. Enstrophy will be injected into the
the domain at rate proportional to the shear within the jet-bow shock
complex.  This can be estimated using a model for the growth of the
bow shock in both the lateral and transverse dimensions.  A number of
authors have provided analytical models for the shape of a jet driven
or clump-driven bow shock \citep{Raga, Cabrit}.  In \cite{Ostriker} a
thin shell model was presented.  In these models the width of the bow
shock, $r$, took the following asymptotic form when the height of the
jet $z$ was much larger than $r_j$
\begin{equation}
z(r) \sim \left( \frac{r}{r_j} \right)^3 (v_s r_j) (3 \beta c_s) \label{zjet}
\end{equation}
Where $v_s$ is the speed of the jet head, $c_s$ is the post
shock sound speed and $\beta$ is a term of order unity reflecting the
effect of cooling on transverse momentum exiting the jet head.  Using
the simple time dependence $z = v_s t$ one finds for that the maximum
bow width $r_{max}$ (at the base of the flow) grows as
\begin{equation}
r_{max} = (3 \beta c_s r_j)^{1/3}t^{1/3}
\label{Rjet}
\end{equation}
A simple estimate of the vorticity magnitude then comes from integrating the
shear in the bow shock over a the bow shock shape.
\begin{eqnarray}
\left| \omega \right| = \frac{\partial v_z}{\partial R} \sim \frac{v_j}{r_{max}}
\label{omega}
\end{eqnarray}
The total magnitude of vorticity in the grid then becomes
\begin{eqnarray}
\left| \Omega \right| = \int \left| \omega \right|dV = 2 \pi \int_{z=0}^{v_s t} \int_{r=0}^{r_{max}} r \left| \omega \right| dr dz
\label{omega1}
\end{eqnarray}
which can be evaluated to
\begin{eqnarray}
\left| \Omega(t) \right| = C t^{4/3}
\end{eqnarray}
Thus for a single jet we expect the total average enstrophy injected
into the grid to increase as a power law $t^n$ in time with $n > 1$.

\begin{figure}[!h]
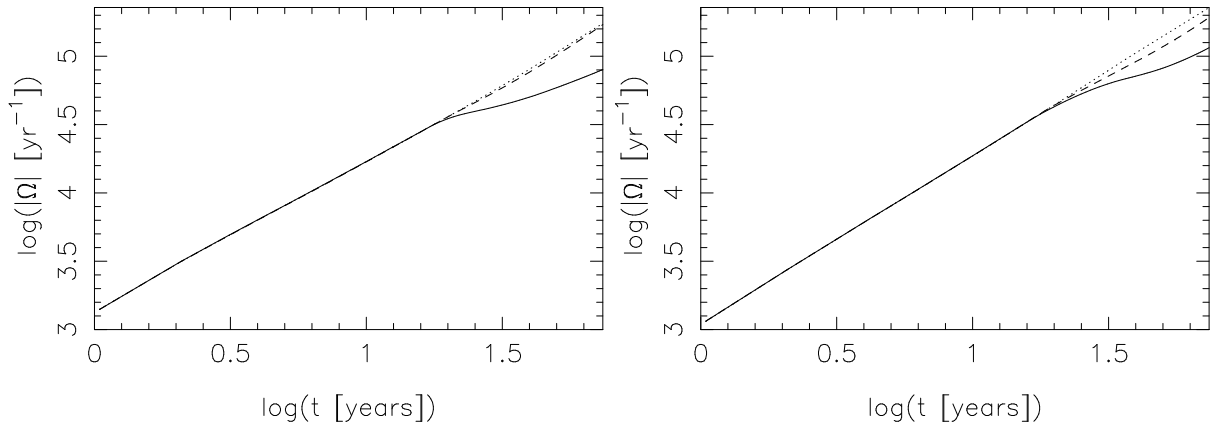

\includegraphics[angle=-90,clip=true,width=0.48\textwidth]{f7a.ps}
\includegraphics[angle=-90,clip=true,width=0.48\textwidth]{f7b.ps}
\caption{The sum of the vorticity magnitude generated in the
simulation domain as a function of time for the collimated (left) and
WAJ (right) simulations.  Solid lines correspond to direct collision,
dashed lines correspond to impact parameter $b=2r_j$, dotted lines
correspond to the non-interacting case. \label{f7}}
\end{figure}

Figure \ref{f7} shows the sum of the magnitude of vorticity generated
in the simulation domain as a function of time.  We have integrated
the enstrophy over the whole domain for each case. First, for the
collimated jet cases, we note that the non-interacting case exhibits a
power law behavior, as derived above, with $n \approx 1.1$.
Consideration of the directly interacting cases shows them to be less
effective at feeding enstrophy into the grid than the non-interacting
case. Thus, the bulk of the enstrophy in an unperturbed jet does arise
from shear across the jet/bow cross section.  Fig \ref{f7} shows that
the non interacting WAJ case generates slightly more enstrophy than
the collimated jets. This effect can be attributed to their larger bow
shock surface area.

\subsection{Radiative Energy Loss}
\begin{figure}[!h]
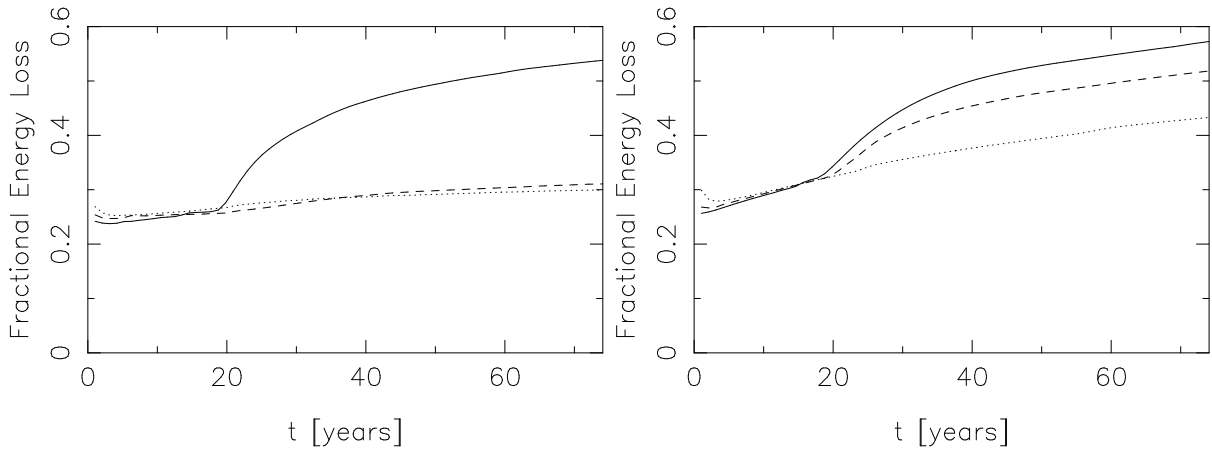

\includegraphics[angle=-90,clip=true,width=0.48\textwidth]{f8a.ps}
\includegraphics[angle=-90,clip=true,width=0.48\textwidth]{f8b.ps}
\caption{Fraction of total energy loss through radiation for the
collimated (left) and WAJ (right) simulations.  Solid lines correspond
to direct collision, dashed lines correspond to impact parameter
$b=2r_j$, dotted lines correspond to the non-interacting
case.\label{f8}}
\end{figure}

We apply the normal adiabatic shock jump conditions at the head of the
bow shock to estimate the radiative energy loss achieved through the
action of a single jet.  The post bow shock temperature $T_{bs}$ is
given in the strong shock limit by $\frac{T_{bs}}{T_{amb}} =
\frac{2\gamma(\gamma-1)}{(\gamma-1)^2}
\left(\frac{v_{bs}}{c_{amb}}\right)^2$ where $c_{amb}=1.3~\textrm{km
s}^{-1}$ is the ambient sound speed and
$v_{bs}=\frac{v_j}{1+(\rho_j/\rho_{amb})^{-1/2}}$ is the estimate of
the bow shock propagation speed of \cite{blondin}.  Taking the
parameters for the flows presented here (table \ref{paramtable}) we
calculate $v_j=140~\textrm{km s}^{-1}$ and
$T_{bs}=7.2\times10^5~\textrm{K}$.  The temperature in the post shock
region falls quickly behind the the bow shock to
$T\sim10^4~\textrm{K}$.  Therefore, most of the thermal energy
generated across the shock front is lost to radiation.  The percentage
of energy injected into the grid that is lost through shock heating
and subsequent cooling within an annulus of cylindrical radius $r_j$
is given by $\frac{E_{rad}}{E_{injected}}=\frac{3 \rho_{bs} R
T_{bs}}{\rho_j v_j^2+3 \rho_j R T_j} \frac{v_{bs}}{v_j}=0.2$ where the
gas density immediately behind the bow shock in the strong shock limit
is $\rho_{bs}=\frac{\gamma+1}{\gamma-1}\rho_j$ and $R$ is the gas
constant.  We emphasize that this calculation includes only the
dominant source of radiative loss at the head of the bow shock and
excludes the radiative losses across the inner wind shock nor does it
include radiative losses along the oblique edges of the bow shock.
This result is therefore a lower-limit prediction of the radiative
energy loss due to the propagation of non-interacting jets.  The
fractional radiative loss achieved in our simulation of
non-interacting jets is $\sim0.25$ (figure \ref{f8}), in agreement
with this result.

The direct collision of outflow streams results in significant
additional energy lost through radiation relative to the
non-interacting case.  The fraction of the total energy budget
invested into the domain that has been emitted as radiation as a
function of time is shown in figure \ref{f8}.  In the collimated jet
case, the direct collision of the outflow streams doubles the
radiative energy loss.  This accounts for more than half of the total
energy imparted to the outflow.  The WAJ cases produce slightly
greater radiative losses than their collimated jet counterparts owing
to the larger working surface at the head of the outflow.  The
increased radiative losses achieved through collisions reduces the
energy budget available for driving turbulent motions after the
driving source has expired and the flow has been subsumed into the
parent cloud.

\subsection{The Adiabatic Case}
\begin{figure}[!h]
\includegraphics[clip=true,width=0.32\textwidth]{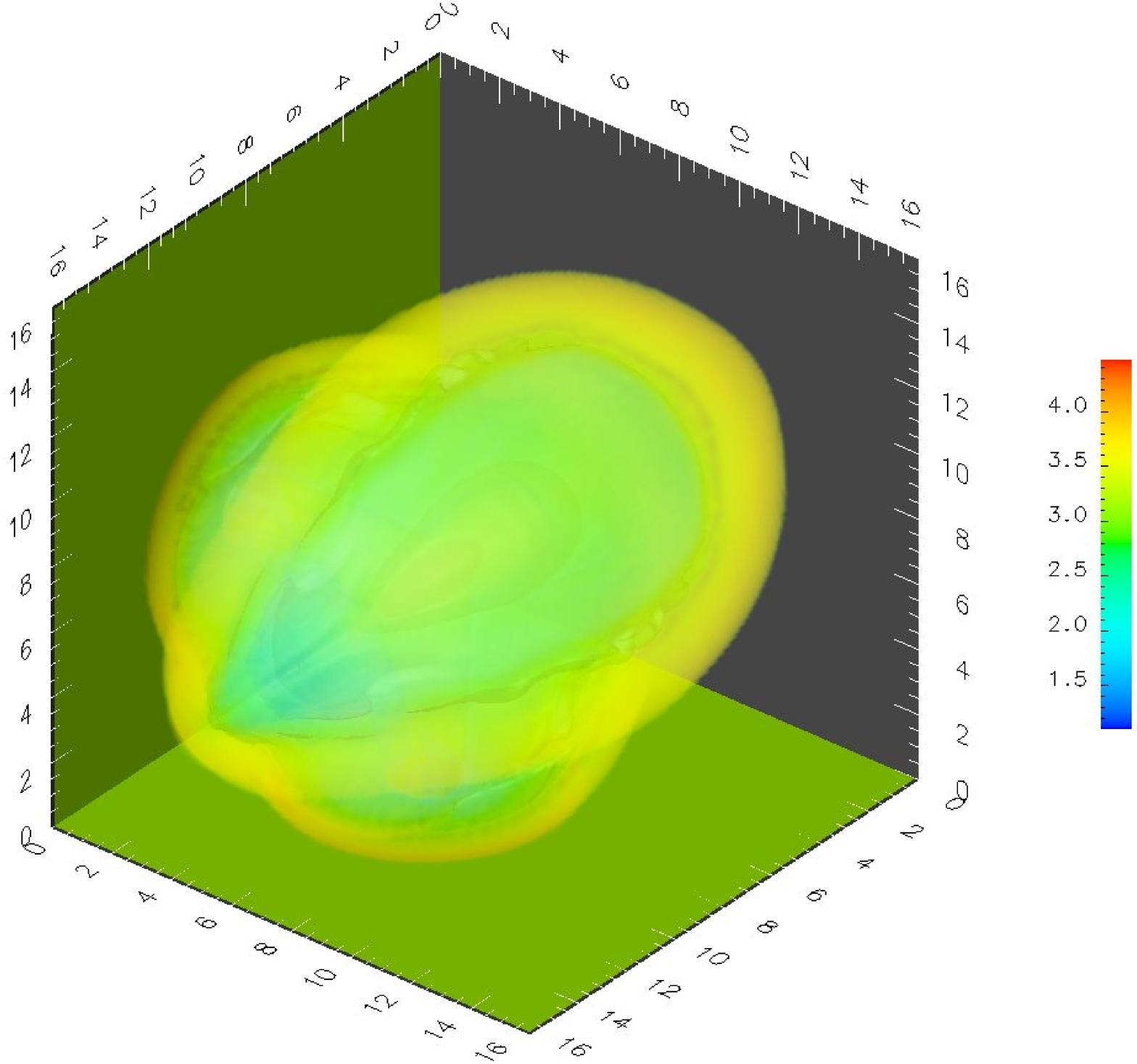}
\includegraphics[clip=true,width=0.32\textwidth]{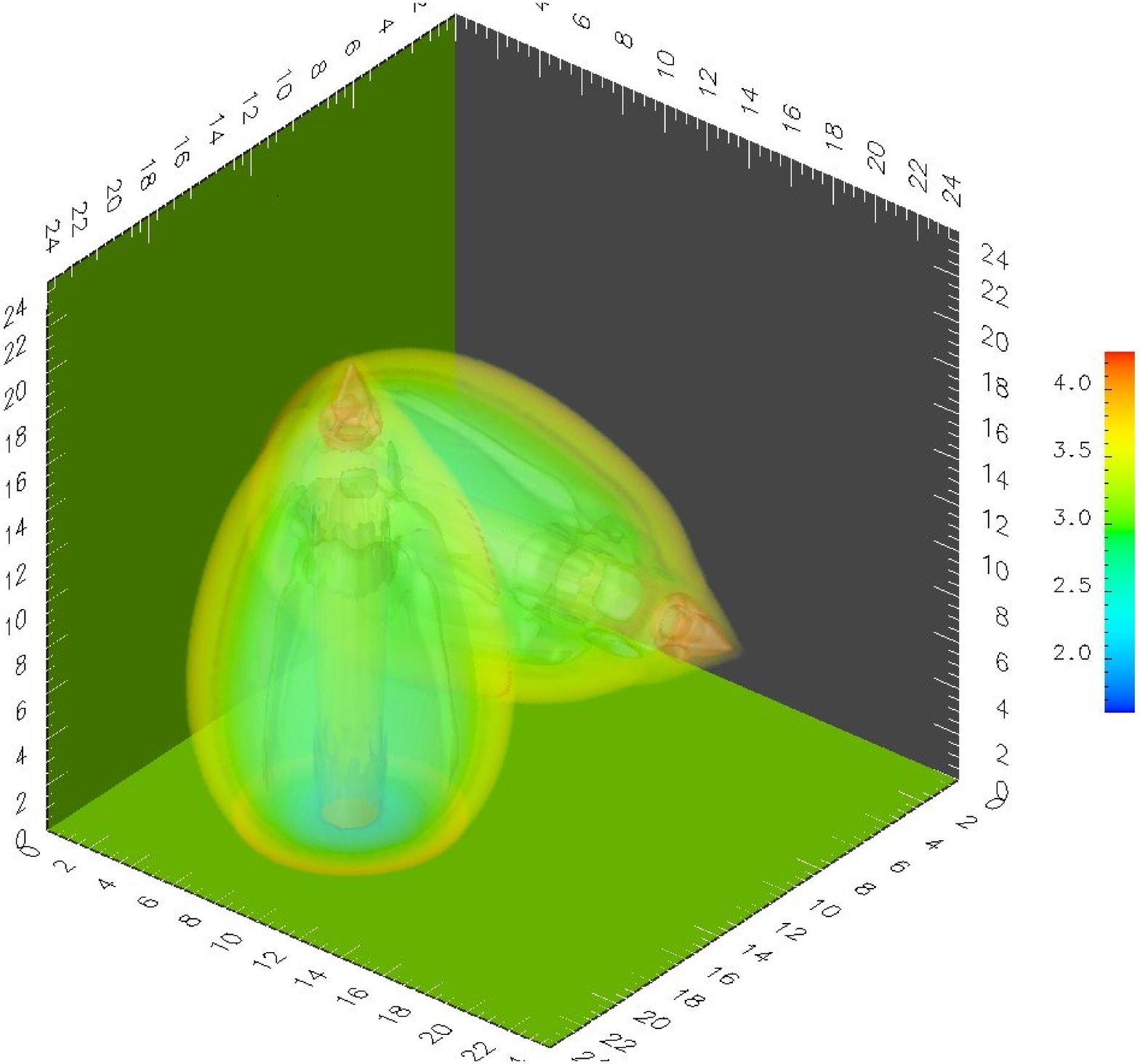}
\caption{A semi-transparent volume rendering of gas density in units
of $\log(\textrm{cm}^{-1})$ at time $t=60~\textrm{yr}$ for the
adiabatic collimated jet simulations.  Semi-transparent isosurfaces
are plotted at $\log(\rho)=2.0, 3.0, 4.0~\textrm{cm}^{-1}$.  The axes
are labeled in units of $r_j=100~\textrm{AU}$.  Impact parameters of
$b=0$ (left), and $b=5.33r_j$ (right) are shown.
\label{f9}}
\end{figure}

To quantify the effect of radiation on outflow collisions we have
performed collimated jet collisions simulations using a polytropic
equation of state for a monotonic gas with impact parameters $b=0$ and
$b=8r_j$ (figure \ref{f9}).  Because the pressure behind the bow shock
is not lost through radiation, the bow shocks are wider than in the
radiative cases.  Note that the flows in the $b=8r_j$ case do interact
slightly.  Approximately 30\% of the injected flows' cross sectional
area overlap at the widest point of the bow shock.  As we have noted
earlier, such grazing collisions have little measurable effect on the
global flow.  The morphology of the resultant flow in the direct
collision case is markedly different from the radiative cases.  The
thermal energy deposited into the flow at the collision shock is
retained and the thermal pressure in this region is dynamically
significant.  Because the thermal pressure drives the flow
isotropically, the resultant flow has a comparable spatial extent in
all three dimensions and is driven over a considerably larger volume
than in the radiative case.

\begin{figure}[!h]
\includegraphics[angle=-90,clip=true,width=0.48\textwidth]{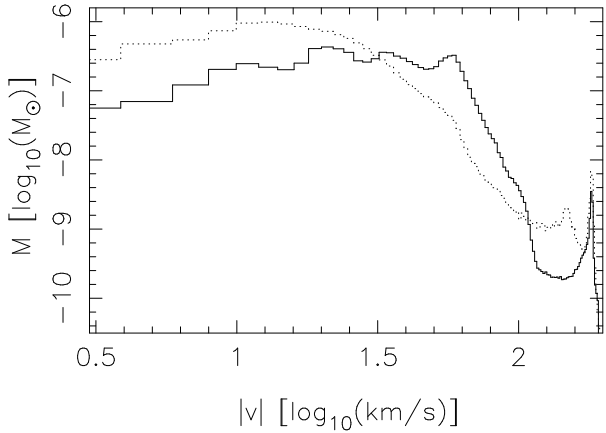}
\caption{Ambient gas mass vs. velocity magnitude histograms at
$t=60~\textrm{yr}$ are shown for the adiabatic collimated jet
simulations.  Solid lines correspond to direct collision and dotted
lines correspond to the non-interacting case. \label{f10}}
\end{figure}

In figure \ref{f10} we show the ambient mass distribution as a
function of the magnitude of the flow velocity at the end of the
adiabatic simulations.  The adiabatic flows reveal a much steeper
power law dependence over a narrower range of velocity $(|v| \sim 50 -
100 \textrm{km s}^{-1})$ than the radiative case.  Comparing the
radiative mass distribution (figure \ref{f6}) to the adiabatic, we
note that the adiabatic collision has resulted in considerably more
ambient material entrained into the flow at lower velocity.  Because
the resultant flow consumes a larger volume, considerably more
material can be entrained into the flow if radiative losses are
suppressed.

\section{Discussion}
\subsection{Result Summary}
The direct collisions have the least potential to drive turbulence in
their environment.  The two injected streams merge into a single
redirected flow across a ``collision shock'' at the vertex of the
interaction region.  The redirected outflow is condensed into a narrow,
high density flow owing to strong radiative cooling.  The gas
compression achieved by the action of the radiative collision shock
allows the redirected flow to remain within a more spatially confined
region than if the flows did not collide.  The surface area of the bow
shock driven by the advance of the redirected flow is less than what
is realized in the absence of collision.  The reduction of the net
surface area of the advancing bow shock is responsible for the reduced
rate of entrainment of ambient material into the flow.  A significant
fraction of the energy of the outflow streams is radiated away in the
post-collision shock region.  The resultant outflow has reduced energy
budget to contribute to the turbulent energy of the region.

The indirect, $b=2r_j$, collisions have little effect on the
outflows' potential to drive turbulence.  While the driving winds
in the WAJ $b=2r_j$ case interact, only the fraction of material
in the intersecting outflow limbs pass through a collision shock
and the driving winds do not merge into a completely redirected
flow.  Grazing collimated jets interact through  the lateral edges
of the bow shock.  In this case, the injected gas streams are not
redirected. Only the cocoon of shocked material between the jet
and bow shocks is disrupted by the collision.  The lateral edges
of the bow shock that participate in the collision contain only a
small fraction of the momentum of the overall outflow which is not
sufficient to produce any significant effect on the entrainment or
deposition of momentum into ambient material.  The vorticity
produced within the shear layer around the working surface at the
head of the outflow is spatially redistributed but not
significantly enhanced through the interaction.

\subsection{Statistics of Global Cloud Support}
In \S\ref{s1}, we estimated the stellar density necessary to achieve
an appreciable number of protostellar outflow collisions.  However,
the results of this work have shown that only low impact parameter
collisions where the high speed, preshock outflow gas streams collide
directly are able to disrupt the global characteristics of the flow.
To estimate the stellar density required to achieve a significant
number of direct collisions, we repeat the calculation setting
$R=200~\textrm{AU}$, characteristic of the radial extent of the
preshock outflow streams.  This yields a critical density for direct
collisions of
$N^{direct}_{critical}=5\times10^6~\textrm{pc}^{-3}$. Such a stellar
density is three orders of magnitude higher than the most dense star
forming regions. Thus the direct collision of preshock outflow
material is unlikely.  While protostellar outflow collisions are likely
common, the direct collision of the driving winds is not.  Outflow
collisions therefore do not influence the turbulent energy budget of
the parent cloud on a global scale.

\section{Conclusions}
The radiative energy losses when unshocked protostellar outflow
streams directly collide reduces the kinetic energy available to
deposit into the molecular cloud as turbulent energy.  The high degree
of compression of outflow gas induced by cooling from such a collision
prevents the redirected outflow from spraying over a large spatial
region.  Furthermore, the collision reduces the redirected outflow's
ability to entrain and impart momentum into the ambient cloud.  The
cooling of the interaction region produces a reduced bow shock surface
area over which outflow momentum can be exchanged with ambient gas.
``Grazing'' collisions, where only the cocoon of shock-decelerated gas
or a small fraction of the high speed outflow gas collide, however,
have little effect on radiative energy loss or the rate of entrainment
of ambient material into the flow.  Because the direct collision of
protostellar outflows is rare, we conclude that such collisions have
little effect on the turbulent energy budget of molecular clouds.

Based on the results of this study we conclude that if turbulence is
energized by outflows it does not occur through collisions of active
outflows. Instead the mechanical energy of an outflow is most likely
supplied to the turbulent motions of the cloud through the action of
{\it fossil} cavities that remain after the driving source of the
outflow has expired.  The role of individual or possibly overlapping
fossil outflows was explored before \citep{Quillen et al 2005}. This
study focused on NGC 1333 and explored the interaction of the cloud
with slowly moving shells which remain after the outflow source has
either shut down or become signifigantly weakened by the decrease in
$\dot{M}_j$. The fossil cavities were shown to carry significant
momentum and can provide the coupling mechanism between outflow and
turbulent motions in the cloud. Using the bow shock radius and outflow
length in \S\ref{s2}, without accounting for collisons, the volume
fill ratio of outflow cavities exceeds unity at stellar density $>
32~\textrm{pc}^{-3}$. We speculate that as the density of protostars
approaches this value, the parent cloud will become subsumed in motion
driven by randomly oriented fossil cavities.  Future work should focus
on the interaction of fossil outflows that were launched at different
times within a turbulent cloud and their overlap to unravel the exact
mechanisms by which this provides a route to sustained turbulence.

\acknowledgments

This paper benefited from discussions with Alice Quillen, John
Bally, Alyssa Goodman, Hector Arce and Mordecai-Mark Mac Low.
We acknowledge support from the Jet Propulsion Laboratory Spizter
Space Telescope theory grant 051080-001, HST theory grant
050292-001, NSF grants 0507519, AST-0406799, AST 00-98442,
AST-0406823, DOE grant DE-F03-02NA00057, NASA grant
ATP04-0000-0016, the Laboratory for Laser Energetics, and the
Center for Computational Research at the University of Buffalo.

\end{document}